\documentclass{emulateapj}

\newcommand{\iso}{{\em ISO}}
\newcommand{\iras}{{\em IRAS}}
\newcommand{\mum}{\ifmmode{\rm \mu m}\else{$\mu$m}\fi}
\newcommand{\figpath}{./}

\begin{document}

\title{Infrared spectral properties of M giants}

\author{
G.~C.~Sloan\altaffilmark{1,2},
C.~Goes\altaffilmark{1},
R.~M.~Ramirez\altaffilmark{1,2},
K.~E.~Kraemer\altaffilmark{3},
\& C.~W.~Engelke\altaffilmark{3}
}
\altaffiltext{1}{Cornell Center for Astrophysics and Planetary Science,
  Cornell University, Ithaca, NY 14853-6801, sloan@isc.astro.cornell.edu}
\altaffiltext{2}{Carl Sagan Institute,
  Cornell University, Ithaca, NY 14853-6801}
\altaffiltext{3}{Institute for Scientific Research, Boston College,
  140 Commonwealth Avenue, Chestnut Hill, MA 02467, USA}

\begin{abstract}

We observed a sample of 20 M giants with the Infrared 
Spectrograph on the {\it Spitzer Space Telescope}.  Most show 
absorption structure at 6.6--6.8~\mum\ which we identify as 
water vapor, and in some cases, the absorption extends from 
6.4~\mum\ into the SiO band at 7.5~\mum.  Variable stars 
show stronger H$_2$O absorption.  While the strength of the 
SiO fundamental at 8~\mum\ increases monotonically from 
spectral class K0 to K5, the dependence on spectral class 
weakens in the M giants.  As with previously 
studied samples, the M giants show considerable scatter in 
SiO band strength within a given spectral class.  All of the 
stars in our sample also show OH band absorption, most 
noticeably in the 14--17~\mum\ region.  The OH bands behave 
much like the SiO bands, increasing in strength in the K 
giants but showing weaker dependence on spectral class in the 
M giants, and with considerable scatter.  An examination of 
the photometric properties reveals that the $V-K$ color may
be a better indicator of molecular band strength than the
spectral class.  The transformation from Tycho colors to
Johnson $B-V$ color is double-valued, and neither $B-V$ nor
$B_T-V_T$ color increases monotonically with spectral class 
in the M giants like they do in the K giants.

\end{abstract}


\section{Introduction} \label{s.intro} 

Red giants dominate the infrared skies \citep{gg71}.  Their
brightness and prevalence have led to their frequent use as
infrared standard stars.  Originally, the spectra of these 
stars were assumed to be blackbodies in the mid-infrared
\citep[e.g.][]{gm75}, but they actually include strong 
molecular absorption bands.  Spectra from the Kuiper Airborne 
Observatory revealed the presence of CO and SiO bands in the 
spectrum of $\alpha$~Tau and several other late-type giants 
\citep{coh92a,coh92b}.  Red giants are also associated with 
mass-loss and dust production, and these molecules are 
precursors to the dust, adding to the importance of 
understanding their spectral properties.

\cite{her02} used spectra from the Short-Wavelength 
Spectrometer (SWS) on the {\it Infrared Space Observatory} 
(\iso) to show that the strength of the CO and SiO bands 
increased with later spectral classes, but with significant
scatter.  The scatter is greater in the M giants than in
the K giants, and it is greater for SiO than CO.  

\citet[][hereafter Paper~I]{slo15} followed up with a study 
of 33 K giants observed by the Infrared Spectrograph 
\citep[IRS;][]{hou04} on the {\it Spitzer Space Telescope} 
\citep{wer04}.  Wavelength coverage limited their study to 
the SiO fundamental at 8~\mum, and for that band, the scatter 
persisted, even though the IRS sample was limited to a 
luminosity class of ``III'', while the earlier SWS sample 
had included bright giants with classes ``II'' and ``IIIa''.  

The spectra of late-type giants can also show absorption from
H$_2$O.  \cite{tsu97} detected H$_2$O at 6.7~\mum\ in SWS 
spectra of giants of spectral class M2 and later.  \cite{tsu01} 
identified several H$_2$O lines at $\sim$6.6~\mum\ in stars 
as early as K5.  \cite{her02} examined a larger sample of SWS 
spectra and found that H$_2$O bands at 6.4--7.0~\mum\ 
commonly appeared in all stars of spectral class $\sim$M2 or 
later.  They were unable to diagnose any further dependencies 
with spectral class.  \cite{ard10} examined H$_2$O absorption 
in a sample observed with the IRS.  They confirmed the 
presence of H$_2$O in M giants, suggesting a turn-on point at 
$\sim$M0, and found little variation in the band strength 
within the M giants.

OH is another absorber in the spectra of late-type giants.
\cite{vm04} identified several bands in the 14--20~\mum\ 
region in spectra from the SWS.  They found that the bands
were stronger than predicted by models, but they were still
too weak for more quantitative conclusions.  The IRS sample
of K giants confirmed that the observed bands were stronger 
than the models.  The sample also shows that the bands 
grow stronger with later spectral class (Paper~I).

To improve our understanding of how the SiO, H$_2$O, and OH
bands behave with spectral class and build on what we have
learned from previous observations of K giants with the IRS 
on {\it Spitzer}, we obtained infrared spectra of 20 M 
giants.  Section~\ref{s.sample} describes the sample and the 
spectra.  Section~\ref{s.analysis} presents our analysis, and
Section~\ref{s.disc} discusses the results.  We summarize
our conclusions in Section~\ref{s.conclude}.
The Appendices detail how we determined the photometric
properties of our sample and also describe our online
spectroscopic data.

\section{The sample} \label{s.sample} 

\subsection{The IRS sample of M giants} \label{s.irs} 

We used several criteria to select the M giant sample 
observed with the IRS.  We aimed to observe at least three 
stars in each spectral class from M0 to M6 in order to 
compare stars both within a given class and from one class 
to the next.  The stars had to have photometry from the {\it 
Infrared Astronomy Satellite} (\iras) at 12 and 25~\mum\ 
consistent with a naked star.  We excluded stars past M6 
because they are usually associated with circumstellar dust.

All targets had to have a luminosity class of ``III'' and not 
classes like ``IIIa'' or ``II--III'', which would indicate 
a difference in luminosity and surface gravity compared to 
the rest of the sample.  And they could not be spectroscopic 
binaries or strong variables.  This last constraint had to
be relaxed somewhat, because some variability is typical at
the latest spectral classes.  Because most M6 giants are
variables and dusty, we chose only two targets with this
spectral class.

\begin{deluxetable*}{llccllrl} 
\tablenum{1}
\tablecolumns{8}
\tablewidth{0pt}
\tablecaption{IRS sample of M giants}
\label{t.irs}
\tablehead{
  \colhead{Target} & \colhead{Alias} & \colhead{RA} & \colhead{Declination} &
  \colhead{Spectral}  & \colhead{Variability} & \colhead{$F_{12}$} & 
  \colhead{AOR key} \\
  \colhead{ }      & \colhead{ }  & \multicolumn{2}{c}{(J2000)\tablenotemark{a}} &
  \colhead{Type\tablenotemark{b}} & \colhead{Class\tablenotemark{c}} & 
  \colhead{(Jy)\tablenotemark{d}} & \colhead{ }
}
\startdata
HD 13570   & \nodata    & 02 10 15.47 & $-$61 05 49.1 & M0 III & \nodata & 0.89 & 21747456 \\ 
HD 19554   & \nodata    & 03 08 09.32 & $-$19 18 10.6 & M0 III & \nodata & 1.31 & 21747712 \\ 
HD 107893  & NSV 19376  & 12 24 03.97 & $-$26 00 50.4 & M0 III & (NSV)   & 1.32 & 21747968 \\ 
\\
HD 17678   & FM Eri     & 02 49 42.59 & $-$17 16 47.3 & M1 III & Lb:     & 1.12 & 21748224 \\ 
BD+47 2949 & HIP 97959  & 19 54 29.30 & $+$47 54 49.7 & M1 III & \nodata & 1.27 & 21748480 \\ 
HD 206503  & \nodata    & 21 45 25.03 & $-$67 06 12.8 & M1 III & \nodata & 0.92 & 21748736 \\ 
\\
HD 122755  & V350 Hya   & 14 04 23.76 & $-$29 53 58.7 & M2 III & Lb      & 1.76 & 21748992 \\ 
HD 177643  & \nodata    & 19 10 54.20 & $-$68 23 59.4 & M2 III & \nodata & 1.03 & 21749248 \\ 
HD 189246  & \nodata    & 20 00 41.66 & $-$40 12 00.8 & M2 III & \nodata & 1.11 & 21749504 \\ 
\\
HD 26231   & CZ Eri     & 04 07 23.80 & $-$39 29 54.9 & M3 III & SRb     & 1.20 & 21749760 \\ 
HD 127693  & \nodata    & 14 33 47.73 & $-$40 01 18.0 & M3 III & \nodata & 0.75 & 21750016 \\ 
HD 223306  & DT Tuc     & 23 48 39.29 & $-$59 03 27.2 & M3 III & Lb:     & 0.72 & 21750272 \\ 
\\
HD 17766   & XX Hor     & 02 48 26.35 & $-$60 24 53.0 & M4 III & Lb:     & 0.87 & 21750528 \\ 
HD 32832   & VX Pic     & 05 03 00.32 & $-$54 05 52.8 & M4 III & SRb     & 1.10 & 21750784 \\ 
HD 46396   & AX Dor     & 06 27 58.88 & $-$66 45 15.9 & M4 III & Lb:     & 1.53 & 21751040 \\ 
\\
HD 68422   & V464 Car   & 08 08 48.91 & $-$61 34 07.6 & M5 III & Lb:     & 1.81 & 21751296 \\ 
HD 74584   & NSV 17963  & 08 41 07.97 & $-$64 36 08.6 & M5 III & (NSV)   & 1.67 & 21751552 \\ 
HD 76386   & CZ Lyn     & 08 57 12.10 & $+$41 20 26.9 & M5 III & SRb     & 2.51 & 21751808 \\ 
\\
HD 8680    & BZ Phe     & 01 24 50.07 & $-$42 45 51.9 & M6 III & Lb:     & 0.88 & 21752064 \\ 
BD+44 2199 & BV CVn     & 12 31 59.43 & $+$43 28 58.8 & M6 III & Lb      & 2.95 & 21752320 \\ 
\enddata
\tablenotetext{a}{Coordinates from \cite{vl07}.}
\tablenotetext{b}{See text for references.}
\tablenotetext{c}{NSV = designated as a new suspected variable in Simbad.}
\tablenotetext{d}{Photometric data are from the IRAS Faint-Source catalog 
  \citep[FSC;][]{fsc} and color corrected by dividing by 1.42.} 
\end{deluxetable*}


Table~\ref{t.irs} presents the sample in order of spectral
class, which are from the Michigan catalogs of spectral 
classifications \citep{mss75,mss78,mss82,mss88}, with three 
exceptions.  \cite{mss78} classified HD 8680 as M3 while 
\cite{jon72} classified it as M6 III; we have adopted the 
latter because it includes a luminosity class.  The first 
complete spectral type for BD+47 2949 appeared in the 1980 
SAO catalog \citep{osc80}.  The classification for BD+44 2199 
is from \cite{upg60}.

In Table~\ref{t.irs}, the fraction of stars identified as
definite variables increases with spectral class.  
\cite{ard10} included one star from each spectral class in 
their {\it Spitzer} Atlas of Stellar Spectra, but they 
generally avoided the more variable sources.  The infrared 
amplitudes are usually a small fraction of the optical
amplitudes, typically one tenth or so \citep{smi04,pri10}.

All 20 M giants were observed by the IRS as part of program
40112, in Cycle 4, between 2007 June and 2008 March.  The
observations used both the Short-Low (SL) and Long-Low (LL)
modules, which obtain spectra with a resolution ($\lambda /
\Delta \lambda$) of $\sim$60--120.  All observations used
the standard IRS nod sequence and integrated for 18 s per nod 
in SL and 28--60 s per nod in LL.  Most of the observations 
were made with an offset peak-up star, which raises the odds 
of slight mispointings.  The spectra were reduced identically 
to the K giants, and Paper~I describes the methods of 
observation and reduction in more detail.

\begin{figure} 
\includegraphics[width=3.4in]{\figpath 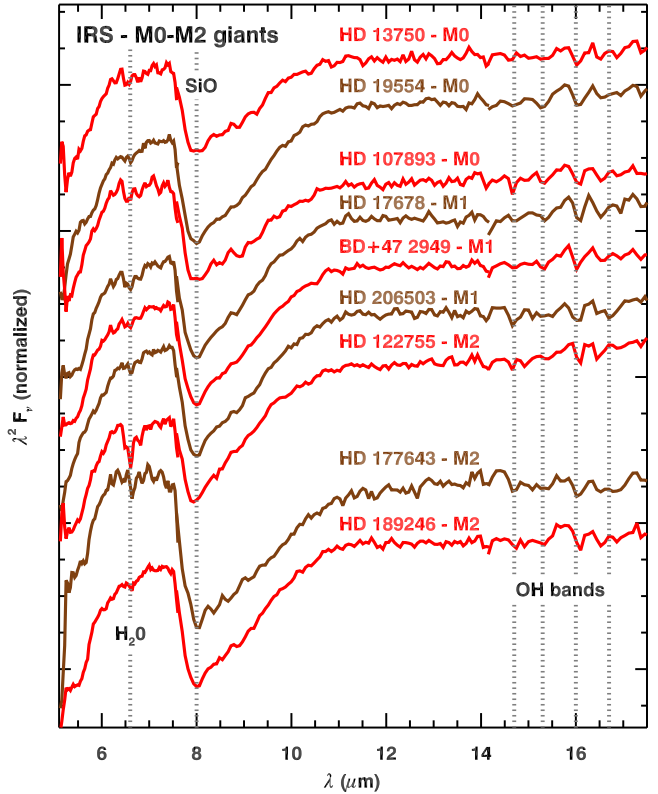} 
\caption{IRS spectra of the early M giants (M0--M2), plotted
in Rayleigh-Jeans units, so that a Rayleigh-Jeans tail would
be a horizontal line.  Vertical dotted lines mark the 
positions of the deepest absorption in the SiO fundamental
at 8~\mum, the H$_2$O absorption at 6.6~\mum, and the 
strongest OH bands from 14.5 to 17~\mum.\label{f.spirs1}}
\end{figure}

\begin{figure} 
\includegraphics[width=3.4in]{\figpath 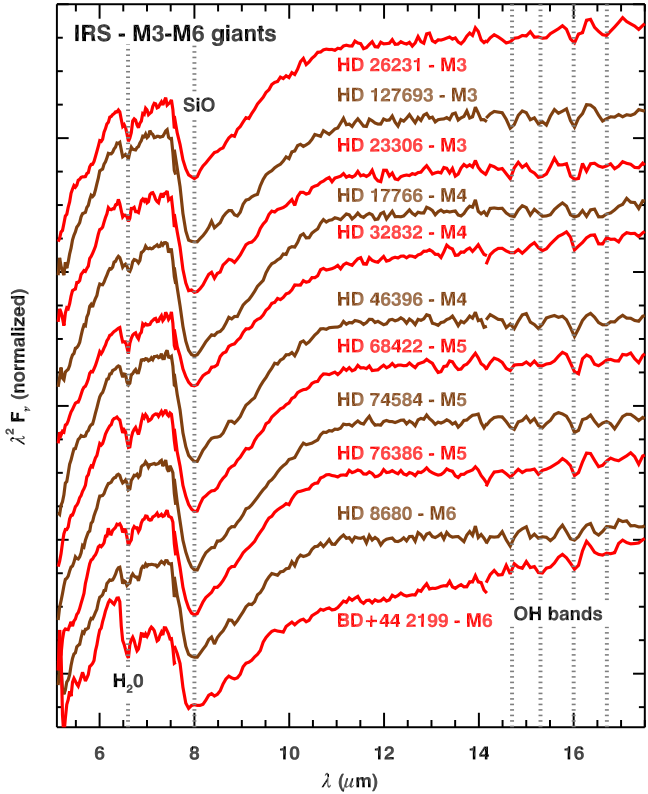} 
\caption{IRS spectra of the late M giants (M3--M6), plotted
and labeled as in Figure~\ref{f.spirs1}.\label{f.spirs2}}
\end{figure}

Figures~\ref{f.spirs1} and \ref{f.spirs2} present the sample 
of M giants observed with the IRS, plotted in Rayleigh-Jeans 
units ($\lambda^2$$F_{\nu}$).  In these units, the 
Rayleigh-Jeans tail of a Planck function would be a 
horizontal line and a typical stellar continuum would rise to 
longer wavelengths, asymptotically approaching the 
Rayleigh-Jeans limit.  

Some of the spectra do not conform to this expected shape, 
most notably HD~177643 (M2) in Figure~\ref{f.spirs1}, with 
the continuum at 6--7.5~\mum\ greater than at 12~\mum\ and 
beyond.  This misbehavior arises when the telescope is 
slightly mispointed, so that the source is not properly 
centered in the slit.  Generally, the slit throughput is a 
function of wavelength and position in the slit due to the 
interactions of the point-spread function (PSF) with the 
edges of the slit.  To first order, mispointed spectra lose
more red flux than blue \citep[e.g.][]{slo12}, but the 
interactions of the Airy rings with the slit edges complicate
the behavior \citep[e.g.][]{slo03a}.  Many of the spectra in 
our sample are affected by small pointing errors, some 
obvious as in the case of HD~177643, and some more subtle.  

All of the spectra show strong absorption bands from
SiO.  Most also show absorption structure at $\sim$6.6~\mum,
which we identify as H$_2$O bands (Section~\ref{s.h2o}).
In some cases, especially in the later M giants, the H$_2$O
absorption appears to extend to the SiO band head at 
7.5~\mum.  In most of the spectra, the OH bands at 
$\sim$14--17~\mum\ are only marginally detected.  

\subsection{Other IRS samples} 

We will also consider the SiO measurements of the sample of
33 K giants from Paper~I.  These stars were selected with
similar criteria to those for our M giants.  Because the K 
giants were observed as part of the IRS calibration program, 
the sample includes more stars in each spectral class.  In
most cases, the K giants were observed at least twice, and 
those stars selected as IRS standards were observed many more 
times.  Each M giant in the current IRS sample was only 
observed once.  For both the K and M giants, variable stars 
were avoided when possible.  While that proved to be a 
challenge for the M giants, the K giants in our sample are
generally non-variable.

Our sample also includes five bright K giants considered by 
Paper~I.  The K2 giant $\xi$~Dra was observed 
repeatedly as a standard for the high-resolution IRS modules.  
The other four were observed to cross-calibrate with previous 
infrared space missions.  These sources were too bright to 
be observed with SL, but they proved useful, both in Paper~I 
and here, for studying the OH bands visible in LL (at 
$\sim$14-18~\mum).  This paper also uses five of the 33 K 
giants from the larger sample in Paper~I as comparision 
sources for OH, because they were observed repeatedly, giving
us high-quality spectra in LL.  

\subsection{SWS sample} \label{s.sws} 

\begin{deluxetable*}{llccllllr} 
\tablenum{2}
\tablecolumns{9}
\tablewidth{0pt}
\tablecaption{SWS sample of K and M stars}
\label{t.sws}
\tablehead{
  \colhead{Target} & \colhead{Alias} & \colhead{RA} & \colhead{Declination} & \colhead{Spectral} & 
  \colhead{Spectral} & \colhead{Variability} & \colhead{SWS} & \colhead{$F_{12}$}  \\
  \colhead{ }      & \colhead{ }  & \multicolumn{2}{c}{(J2000)\tablenotemark{a}} & \colhead{Type} &
  \colhead{Reference\tablenotemark{b}} & \colhead{Class} & 
  \colhead{Source\tablenotemark{c}} & \colhead{(Jy)\tablenotemark{d}}
}
\startdata
$\alpha$ Boo   & HR 5340  & 14 15 39.67 & $+$19 10 56.7 & K1.5 III      & M38, K89   & \nodata & E06     & 558.5 \\ 
$\gamma^1$ And & HR 603   & 02 03 53.92 & $+$42 19 47.4 & K2 III        & R52, Bi54  & \nodata & E06 (t) &  69.4 \\ 
$\alpha$ Ari   & HR 617   & 02 07 10.41 & $+$23 27 44.7 & K2 III SB     & R52, S52   & (NSV)   & E06 (t) &  54.8 \\ 
$\xi$ Dra      & HR 6688  & 17 53 31.73 & $+$56 52 21.5 & K2 III        & R52, M53   & \nodata & E06 (t) &  11.9 \\ 
\\
$\sigma$ Oph   & HR 6498  & 17 26 30.88 & $+$04 08 25.3 & K3 II var     & R52        & (NSV)   & E06 (t) &  13.2 \\ 
$\lambda$ Gru  & HR 8411  & 22 06 06.89 & $-$39 32 36.1 & K3 III        & MC2        & \nodata & E06 (t) &   8.2 \\ 
$\alpha$ Tuc   & HR 8502  & 22 18 30.09 & $-$60 15 34.5 & K3 III SB     & B62        & \nodata & E06 (t) &  41.8 \\ 
$\beta$ UMi    & HR 5563  & 14 50 42.33 & $+$74 09 19.8 & K4 III var    & R52, M53   & (NSV)   & E06     & 112.9 \\ 
\\
$\delta$ Psc   & HR 224   & 00 48 40.94 & $+$07 35 06.3 & K5 III        & N47, R52   & \nodata & E06 (t) &  14.6 \\ 
$\gamma$ Phe   & HR 429   & 01 28 21.93 & $-$43 19 05.7 & K5 Ib         & S59        & \nodata & E06 (t) &  49.1 \\ 
$\alpha$ Tau   & HR 1457  & 04 35 55.24 & $+$16 30 33.5 & K5 III        & M43, R52   & Lb:     & E06     & 492.7 \\ 
H Sco          & HR 6166  & 16 36 22.47 & $-$35 15 19.2 & K5 III        & MC3        & (NSV)   & E06 (t) &  23.6 \\ 
$\gamma$ Dra   & HR 6705  & 17 56 36.37 & $+$51 29 20.0 & K5 III        & M43, R52   & \nodata & E06     & 109.2 \\ 
\\
$\beta$ And    & HR 337   & 01 09 43.92 & $+$35 37 14.0 & M0 III var    & M43        & (NSV)   & E06     & 201.9 \\ 
$\mu$ UMa      & HR 4069  & 10 22 19.74 & $+$41 29 58.3 & M0 III SB     & Bi54, E55  & (NSV)   & E06 (t) &  71.1 \\ 
7 Cet          & HR 48    & 00 14 38.42 & $-$18 55 58.3 & M1 III        & MC4, A67   & Lb:     & E06 (t) &  30.2 \\ 
$\delta$ Oph   & HR 6056  & 16 14 20.74 & $-$03 41 39.6 & M1 III        & Bi54, E55  & (NSV)   & E06 (t) & 105.4 \\ 
$\alpha$ Cet   & HR 911   & 03 02 16.77 & $+$04 05 23.1 & M2 III        & Bl54, Bi54 & Lb:     & E06     & 165.3 \\ 
$\beta$ Peg    & HR 8775  & 23 03 46.46 & $+$28 04 58.0 & M2 II-III var & Bl54, Bi54 & Lb      & E06     & 272.7 \\ 
\\
$\rho$ Per     & HR 921   & 03 05 10.59 & $+$38 50 25.0 & M3 III var    & K42        & SRb     & E06     & 217.3 \\ 
$\pi$ Aur      & HR 2091  & 05 59 56.10 & $+$45 56 12.2 & M3 II var     & M43, Bi54  & Lc      & E06 (t) &  75.8 \\ 
$\delta$ Vir   & HR 4910  & 12 55 36.21 & $+$03 23 50.9 & M3 III        & W57. H58   & (NSV)   & E06 (t) & 114.4 \\ 
$\beta$ Gru    & HR 8636  & 22 42 40.05 & $-$46 53 04.5 & M3 II var     & V56, E60   & Lc:     & E06     & 663.5 \\ 
\\
$\gamma$ Cru   & HR 4763  & 12 31 09.96 & $-$57 06 47.6 & M4 III        & MC1        & (NSV)   & E06     & 609.4 \\ 
$\delta^2$ Lyr & HR 7139  & 18 54 30.28 & $+$36 53 55.0 & M4 II         & M73        & SRc:    & E06     & 109.7 \\ 
57 Peg         & HR 8815  & 23 09 31.46 & $+$08 40 37.8 & M4S           & K54        & SRa     & E06 (t) &  57.0 \\ 
TU CVn         & HR 4909  & 12 54 56.52 & $+$47 11 48.2 & M5 III var    & U60        & SRb     & S03     &  40.4 \\ 
2 Cen          & HR 5192  & 13 49 26.72 & $-$34 27 02.8 & M5 III        & MC3        & SRb     & S03     & 179.9 \\ 
R Lyr          & HR 7157  & 18 55 20.10 & $+$43 56 45.9 & M5 III var    & K45, E57   & SRb     & S03     & 261.1 \\ 
\\
$\rho^1$ Ari   & HR 867   & 02 55 48.50 & $+$18 19 53.9 & M6 III var    & Bl54       & SRb     & S03     & 103.7 \\ 
V537 Car       & HD 98434 & 11 18 43.74 & $-$58 11 11.1 & M6 III        & MC1        & SRb     & S03     &  35.8 \\ 
OP Her         & HR 6702  & 17 56 48.53 & $+$45 21 03.1 & M6S           & K54        & SRb     & S03     &  38.1 \\ 
NU Pav         & HR 7625  & 20 01 44.75 & $-$59 22 33.2 & M6 III        & MC1        & SRb     & S03     & 163.0 \\ 
\enddata
\tablenotetext{a}{Coordinates from \cite{vl07}.}
\tablenotetext{b}{References are meant to be representative:  
  A67 \citep{app67}, B62 \citep{bus62}, Bi54 \citep{bid54}, 
  Bl54 \citep{bla54}, E55 \citep{egg55}, E57 \citep{egg57}, 
  E60 \citep{egg60}, H58 \citep{hw58},
  K42 \citep{kee42}, K45 \citep{kh45}, 
  K54 \citep{kee54}, K89 \citep{km89}, M38 \citep{mor38}, 
  M43 \citep{mkk43}, M53 \citep{mhj53}, M73 \citep{mk73}, 
  MC1--4 are the Michigan catalogue, vol.\ 1--4 
  \citep{mss75,mss78,mss82,mss88}, N47 \citep{nva47}, 
  R52 \citep{rom52}, S52 \citep{sha52}, S59 \citep{sto59}, 
  U60 \citep{upg60}, V56 \citep{dev56}, W57 \citep{wb57}.}
\tablenotetext{c}{Short-Wavelength Spectrometer data are from:  
  E06 \citep{eng06}, S03 \citep{slo03b}.  Spectra marked ``(t)'' 
  are based on a template to the red of 12~\mum\ and are virtually 
  identical at those wavelengths.}
\tablenotetext{d}{Photometric data arefrom the IRAS 
  Point-Source catalog \citep[PSC;][]{psc} and color 
  corrected by dividing by 1.42.} 
\end{deluxetable*}

We have also analyzed a sample of spectra from the SWS on 
\iso, selected based on the catalog of infrared spectral
classifications of the SWS database by \cite{kra02}.  They
classified spectra of naked stars with oxygen-rich absorption
bands as ``1.NO''.  We started with the 48 1.NO sources and
dropped 15 because the spectra were too noisy, too dusty, or
had other flaws.  The remaining 33 targets included 26
spectra re-processed by \cite{eng06}.  Among the improvements
to previous versions of the SWS data, they modified the shape
of the spectra to force them to agree with photometric 
measurements in the near- and mid-infrared.  Their sample did 
not include any sources with spectral classes later than M4.  
The remaining seven spectra in our sample, all M5 or M6, are 
from the SWS atlas of \cite{slo03b}.  These spectra have
resolutions which vary between $\sim$250 and 600.

The SWS sample includes one supergiant ($\gamma$~Phe, K5~Ib), 
two MS stars, and four bright giants (luminosity class II).
The infrared spectral properties of these sources do not
stand out in any significant way from the rest of the sample.

\begin{figure} 
\includegraphics[width=3.4in]{\figpath 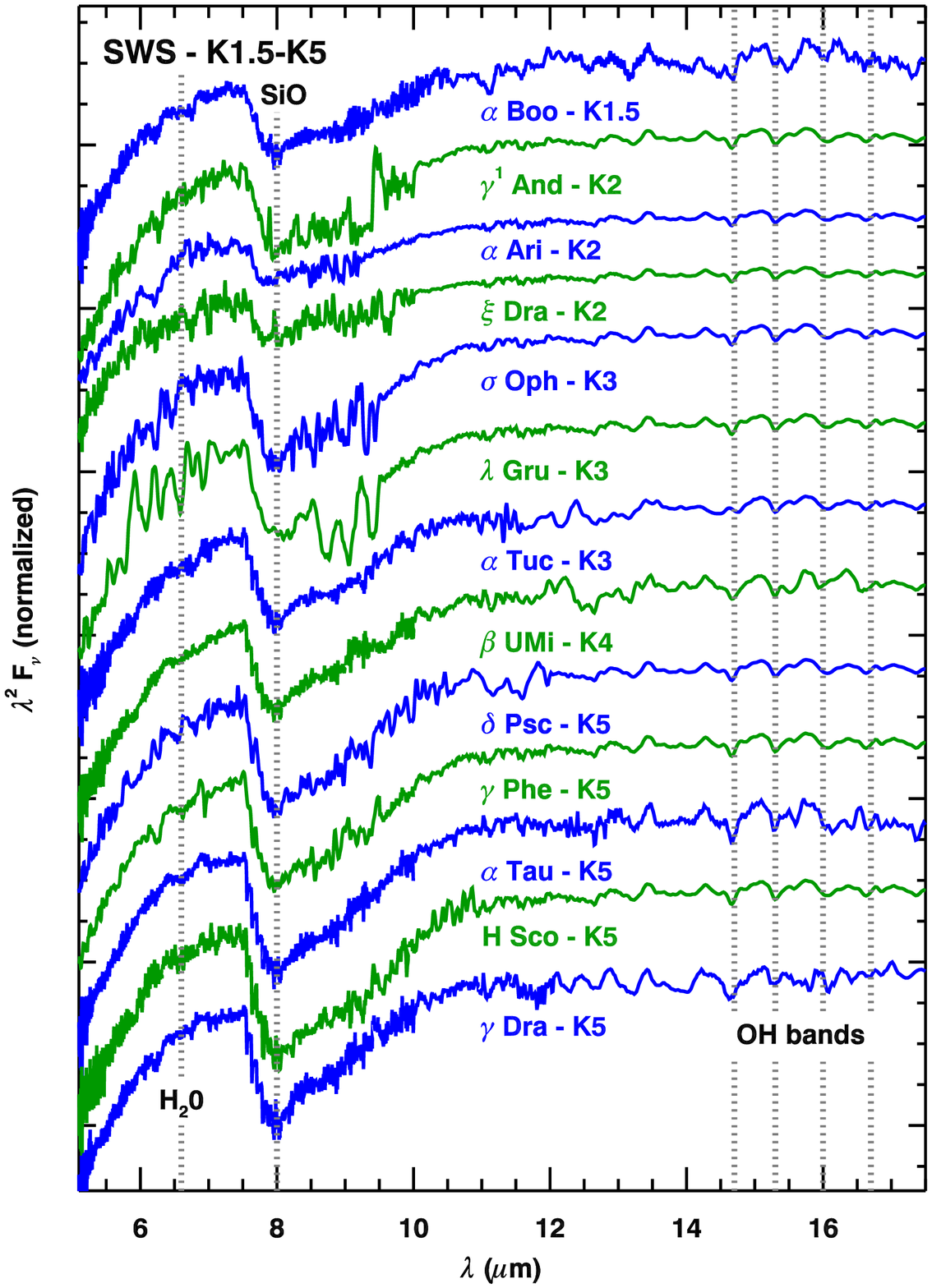} 
\caption{SWS spectra of the K giants and supergiants showing 
oxygen-rich molecular absorption, plotted and labeled as in 
Figure~\ref{f.spirs1}.  These spectra were processed and 
calibrated by \cite{eng06}.\label{f.spsws1}}
\end{figure}

\begin{figure} 
\includegraphics[width=3.4in]{\figpath 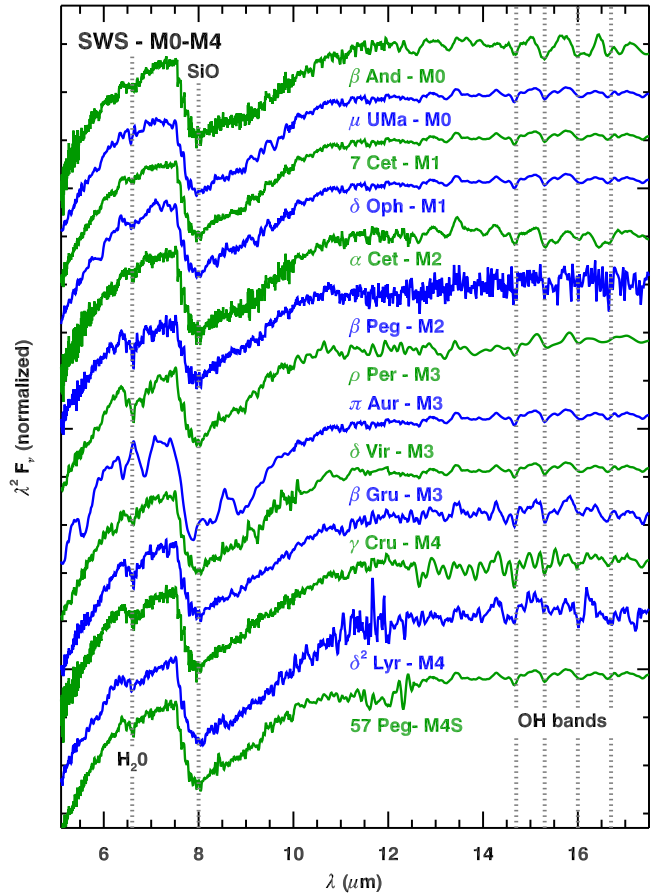} 
\caption{SWS spectra of the early M giants (M0--M4), plotted
and labeled as in Figure~\ref{f.spirs1}.  These spectra are 
from \cite{eng06}.\label{f.spsws2}}
\end{figure}

Figure~\ref{f.spsws1} shows the SWS spectra of the 12 K
giants and one K supergiant in our comparison sample.  The
spectra show a SiO band which grows stronger with later
spectral class.  Three spectra, $\gamma^1$~And, 
$\lambda$~Gru, and $\pi$~Aur, are affected by artificial 
structure in the SiO band, but this has little impact on our 
measured equivalent width.  H$_2$O absorption at 6.6~\mum\ is 
not apparent in any of the spectra.  

Table~\ref{t.sws} notes that in 15 of the 26 SWS spectra
calibrated by \cite{eng06}, the data past 12~\mum\ are based
on a template.  The long-wavelength data in these spectra 
were too noisy and difficult to calibrate, leading 
\cite{eng06} to replace them with the average of all 15, 
fitted to the photometry for each source.  Therefore, while
many of the spectra in Figure~\ref{f.spsws1} show clear OH
band structure, those data are not independent, and we
cannot use them in our analysis.  The same is true in 
Figure~\ref{f.spsws2}.  Consequently, we will limit our 
spectral analysis of the SWS data to the SiO and H$_2$O bands 
and not consider the OH bands.

Figure~\ref{f.spsws2} continues the SWS sequence to M4, 
showing the onset of H$_2$O absorption at spectral class M0
and the continued strengthening of the SiO band.  Again, in
most of the spectra, the OH bands are actually from a 
template spectrum.

\begin{figure} 
\includegraphics[width=3.4in]{\figpath 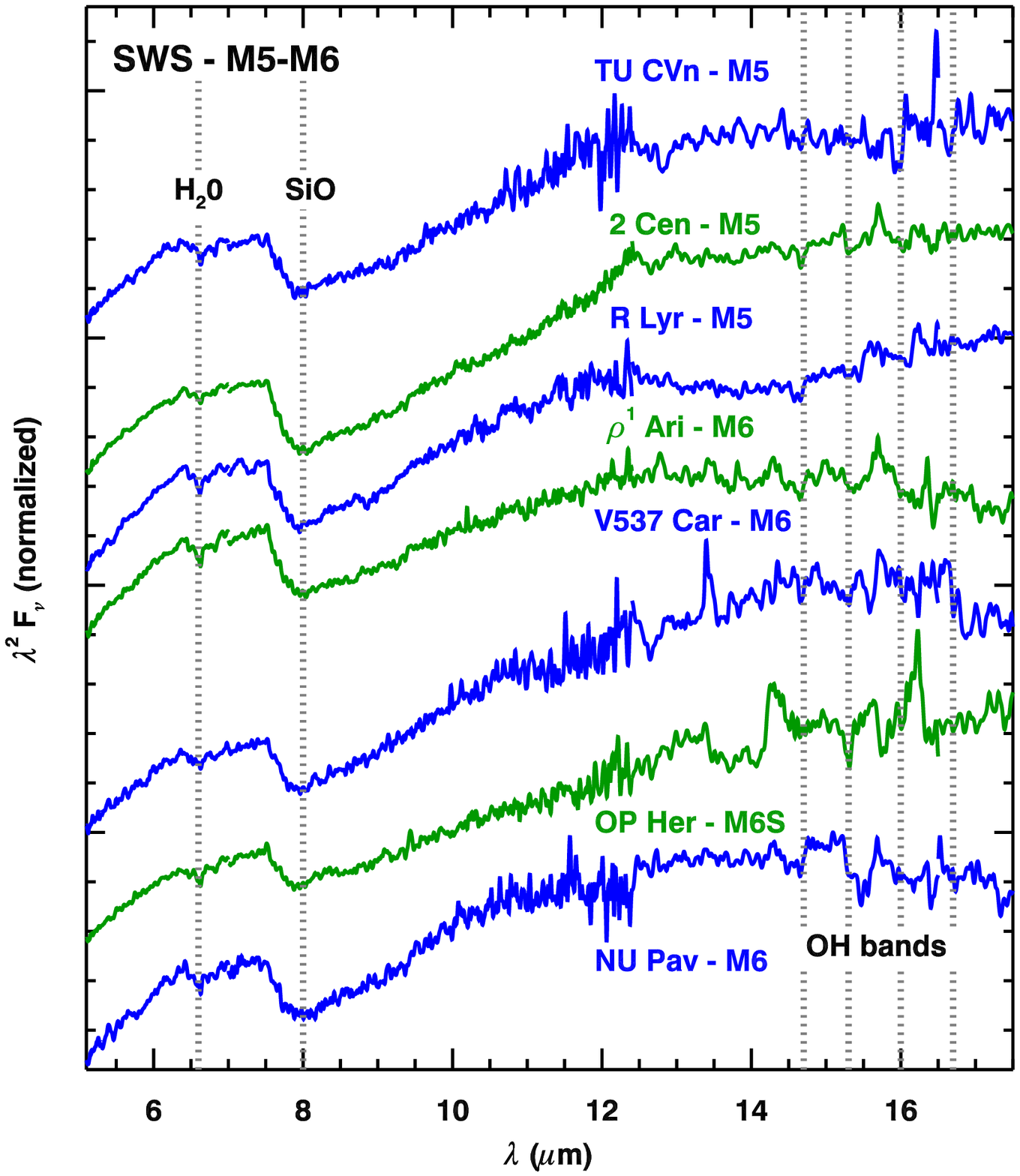} 
\caption{SWS spectra of the late M giants (M5--M6), plotted
and labeled as in Figure~\ref{f.spirs1}.  These spectra are 
from \cite{slo03b}.\label{f.spsws3}}
\end{figure}

Figure~\ref{f.spsws3} plots the seven spectra from the SWS 
not re-processed by \cite{eng06}.  As a group, these spectra
show more artifacts, especially at $\sim$11--12 and 
14--18~\mum.  The spectra are also redder than the 
earlier-type sources, suggesting the presence of optically 
thin dust emission.  The spectra of 2~Cen, R~Lyr, 
$\rho^1$~Ari, and to a lesser degree NU~Pav all have 
inflections at $\sim$11--12~\mum, which are consistent with
the presence of alumina dust.  

Because \cite{eng06} corrected the shape of the spectra and
forced them to match the photometry, the overall shape of 
the continuum should be reasonably reliable in
Figures~\ref{f.spsws1} and \ref{f.spsws2}.  The general
shapes of the spectra in Figure~\ref{f.spsws3}, however, 
are strongly affected by dust and may also be affected by 
uncorrected artifacts in the data.

\section{Analysis} \label{s.analysis} 

\subsection{Measuring the SiO and H$_2$O bands} 

Paper~I described the general procedure for measuring
the strength of the SiO band at 8~\mum.  They simultaneously
fitted an Engelke function \citep{eng92}, a template SiO
profile, and an interstellar extinction profile based on the 
local extinction spectrum of \cite{ct06}.  The Engelke 
function mimics the effect of the H$^{-}$ ion, which has an
opacity that increases with wavelength, by decreasing the
effective blackbody temperature to longer wavelengths.  The
template SiO profile is an average of the SiO band in 
several of the SWS spectra, as described in Paper~I.  The
fitting method forces the Engelke function through the 
IRS data at $\sim$6.5--7.5~\mum\ and $\sim$12~\mum, and 
adjusts the temperature, the depth of the SiO band, and the 
interstellar extinction $A_V$ to minimize the $\chi^2$ error 
between the resulting spectrum and the data from 6.8 to 
11.2~\mum.  

The M giant sample forced several modifications to the 
approach in Paper~I.  With the cooler M giants, changes to 
the stellar temperature and $A_V$ were largely degenerate, 
requiring an independent estimate of $A_V$.  We used the 
three-dimensional extinction model of \citet[][hereafter 
D03]{dri03}.  Their software provides two extinctions, and we 
used the rescaled values.  The input distances were based on 
the Hipparcos parallaxes \citep{vl07}.  In the six cases where 
the D03 code did not provide an extinction, we used the 
extinctions for that line of sight by \citet[][hereafter 
SF11]{sf11}.  In seven cases, the D03 extinctions exceeded 
the SF11 extinctions, but the latter should be an upper limit 
since they represent the total extinction along a given line 
of sight to an infinite distance.  In those cases, we used 
SF11.

For the SWS sample, D03 provided $A_V$ estimates for 22 
sources.  For the remaining 11, we used $A_K$ estimates by
\cite{tab09}, converting to $A_V$ using the extinction law
calibrated by \cite{rl85}.  In seven additional cases, we
used the extinctions from \cite{tab09} in place of the D03
values because they were larger than D03.

Most of the M giants observed by both the IRS and SWS exhibit 
absorption from water vapor at $\sim$6.7~\mum.  In some 
spectra, this absorption is more significant, taking a 
substantial notch out of the spectrum from $\sim$6.4~\mum\ 
all the way to the beginning of the SiO band at 7.5~\mum.  
These spectra required modifications to how we fitted the 
continuum.  Instead of minimizing the $\chi^2$ error, we 
simply chose a temperature that forced an Engelke function 
fitted to the spectrum at $\sim$10.5--11~\mum\ to pass 
through the spectrum somewhere between $\sim$6.3 and 
7.5~\mum.  These wavelength ranges could vary from one 
spectrum to the next, depending on the strength of red wing 
of the SiO band, possible contamination from cool dust, and 
the strength and structure of the H$_2$O band.  

\begin{figure} 
\includegraphics[width=3.4in]{\figpath 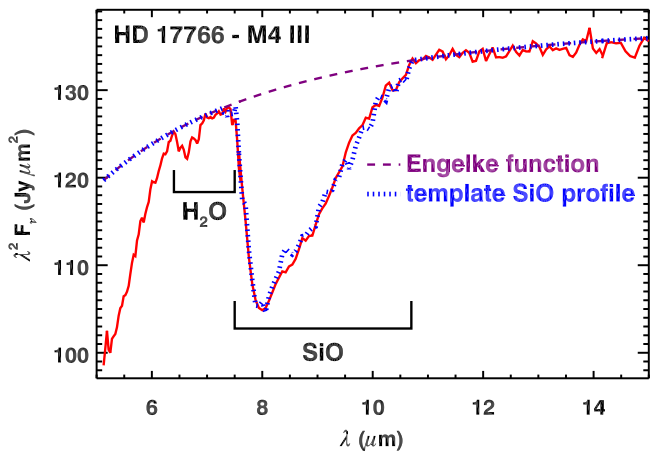} 
\caption{To measure the equivalent width of the SiO and 
H$_2$O bands, we fit an Engelke function, assuming that
the data past $\sim$11~\mum\ and at least some of the
data in the 6--7.5~\mum\ range are from the continuum.  
The equivalent width is integrated underneath the 
continuum, assuming that the absorption from 6.4 to 
7.5~\mum\ is from H$_2$O and from 7.5 to 10.7~\mum\ is 
from SiO.\label{f.exsio}}
\end{figure}

\begin{deluxetable*}{lcclrcc} 
\tablenum{3}
\tablecolumns{7}
\tablewidth{0pt}
\tablecaption{SiO and H$_2$O equivalent widths in the IRS sample}
\label{t.irsew}
\tablehead{
  \colhead{Target} & \colhead{Parallax} & \colhead{$A_V$} & \colhead{$A_V$} & 
  \colhead{Fitted} & \colhead{Eq.\ Width (SiO)} & \colhead{Eq.\ Width (H$_2$O)} \\
  \colhead{ } & \colhead{(mas)} & \colhead{(mag)} & \colhead{Ref.\tablenotemark{a}} &
  \colhead{$T$ (K)} & \colhead{(\mum)} & \colhead{(nm)}
}
\startdata
HD 13570       & ~~~1.3 $\pm$ 0.5 & 0.082 & SF11 &  6350 & 0.194 $\pm$ 0.007 & ~~~ 6.0 $\pm$  3.0 \\
HD 19554       & ~~~2.9 $\pm$ 0.9 & 0.079 & SF11 &  3300 & 0.264 $\pm$ 0.003 & ~~~ 4.8 $\pm$  3.2 \\
HD 107893      & ~~~1.4 $\pm$ 0.8 & 0.108 & D03  & 10000 & 0.218 $\pm$ 0.006 & ~~~ 5.9 $\pm$  2.8 \\
\\
HD 17678       & ~~~1.7 $\pm$ 1.3 & 0.062 & SF11 &  3000 & 0.242 $\pm$ 0.003 & ~~~11.6 $\pm$  2.7 \\
BD+47 2949     & ~~~2.9 $\pm$ 0.6 & 0.202 & D03  &  2950 & 0.238 $\pm$ 0.007 & ~~~ 3.1 $\pm$  5.7 \\
HD 206503      & ~~~1.1 $\pm$ 0.7 & 0.102 & SF11 &  3400 & 0.254 $\pm$ 0.005 & ~~~ 5.7 $\pm$  3.0 \\
\\
HD 122755      & ~~~1.7 $\pm$ 0.9 & 0.110 & D03  &  2850 & 0.220 $\pm$ 0.004 & ~~~31.6 $\pm$  2.3 \\
HD 177643      & ~~~1.3 $\pm$ 0.9 & 0.181 & D03  & 10000 & 0.345 $\pm$ 0.006 & $-$13.1 $\pm$ 17.4 \\
HD 189246      & ~~~1.5 $\pm$ 1.0 & 0.221 & SF11 &  3800 & 0.289 $\pm$ 0.007 & ~~~ 5.5 $\pm$  1.9 \\
\\
HD 26231       & ~~~1.3 $\pm$ 0.7 & 0.025 & SF11 &  2550 & 0.245 $\pm$ 0.003 & ~~~32.1 $\pm$  5.5 \\
HD 127693      & ~~~1.4 $\pm$ 1.1 & 0.233 & D03  &  4850 & 0.292 $\pm$ 0.011 & ~~~ 8.7 $\pm$  2.3 \\
HD 223306      & ~~~2.5 $\pm$ 1.1 & 0.035 & SF11 &  4300 & 0.259 $\pm$ 0.005 & ~~~ 8.1 $\pm$  3.1 \\
\\
HD 17766       & ~~~1.8 $\pm$ 0.8 & 0.060 & SF11 &  3400 & 0.306 $\pm$ 0.003 & ~~~ 7.4 $\pm$  3.4 \\
HD 32832       & $-$0.1 $\pm$ 0.7 & 0.048 & SF11 &  2450 & 0.225 $\pm$ 0.006 & ~~~21.0 $\pm$  2.2 \\
HD 46396       & ~~~1.2 $\pm$ 0.6 & 0.150 & SF11 &  3100 & 0.282 $\pm$ 0.006 & ~~~ 6.9 $\pm$  3.1 \\
\\
HD 68422       & ~~~1.1 $\pm$ 0.9 & 0.381 & D03  &  2850 & 0.281 $\pm$ 0.011 & ~~~27.1 $\pm$  5.3 \\
HD 74584       & ~~~3.3 $\pm$ 0.5 & 0.428 & SF11 &  2750 & 0.291 $\pm$ 0.012 & ~~~11.8 $\pm$  3.8 \\
HD 76386       & ~~~1.8 $\pm$ 0.8 & 0.070 & SF11 &  3000 & 0.273 $\pm$ 0.003 & ~~~15.5 $\pm$  4.2 \\
\\
HD 8680        & ~~~4.5 $\pm$ 1.1 & 0.041 & SF11 &  4200 & 0.259 $\pm$ 0.006 & ~~~13.3 $\pm$  3.8 \\
BD+44 2199     & $-$0.5 $\pm$ 1.1 & 0.031 & D03  &  7950 & 0.323 $\pm$ 0.005 & ~~~67.4 $\pm$  2.6 \\
\enddata
\tablenotetext{a}{References for extinction:  D03 
  \citep{dri03}, SF11 \citep{sf11}.}
\end{deluxetable*}

\begin{deluxetable*}{lrclrcr} 
\tablenum{4}
\tablecolumns{7}
\tablewidth{0pt}
\tablecaption{SiO and H$_2$O equivalent widths in the SWS sample}
\label{t.swsew}
\tablehead{
  \colhead{Target} & \colhead{Parallax} & \colhead{$A_V$} & \colhead{$A_V$} &
  \colhead{Fitted} & \colhead{Eq.\ Width (SiO)} & \colhead{Eq.\ Width (H$_2$O)} \\
  \colhead{ } & \colhead{(mas)} & \colhead{(mag)} & \colhead{Ref.\tablenotemark{a}} &
  \colhead{$T$ (K)} & \colhead{(\mum)} & \colhead{(nm)}
}
\startdata
$\alpha$ Boo   & 88.8 $\pm$ 0.5 & 0.002 & D03  &  4950 & 0.147 $\pm$ 0.000 & \nodata \\
$\gamma^1$ And &  8.3 $\pm$ 1.0 & 0.034 & D03  &  4000 & 0.188 $\pm$ 0.002 & \nodata \\
$\alpha$ Ari   & 49.6 $\pm$ 0.2 & 0.027 & D03  &  5250 & 0.102 $\pm$ 0.001 & \nodata \\
$\xi$ Dra      & 29.0 $\pm$ 0.1 & 0.018 & T09  &  4450 & 0.086 $\pm$ 0.001 & \nodata \\
\\
$\sigma$ Oph   &  3.6 $\pm$ 0.3 & 0.188 & T09  &  3400 & 0.199 $\pm$ 0.007 & \nodata \\
$\lambda$ Gru  & 13.5 $\pm$ 0.2 & 0.042 & D03  &  4600 & 0.220 $\pm$ 0.003 & \nodata \\
$\alpha$ Tuc   & 16.3 $\pm$ 0.6 & 0.083 & D03  &  4900 & 0.164 $\pm$ 0.003 & \nodata \\
$\beta$ UMi    & 24.9 $\pm$ 0.1 & 0.014 & D03  &  3650 & 0.172 $\pm$ 0.000 & \nodata \\
\\
$\delta$ Psc   & 10.5 $\pm$ 0.2 & 0.080 & T09  &  4500 & 0.208 $\pm$ 0.003 & \nodata \\
$\gamma$ Phe   & 14.0 $\pm$ 0.3 & 0.033 & D03  &  3750 & 0.239 $\pm$ 0.001 & \nodata \\
$\alpha$ Tau   & 48.9 $\pm$ 0.8 & 0.063 & T09  &  3850 & 0.267 $\pm$ 0.002 & \nodata \\
H Sco          &  9.5 $\pm$ 0.2 & 0.143 & T09  &  3400 & 0.289 $\pm$ 0.005 & \nodata \\
$\gamma$ Dra   & 21.1 $\pm$ 0.1 & 0.018 & T09  &  4250 & 0.211 $\pm$ 0.001 & \nodata \\
\\
$\beta$ And    & 16.5 $\pm$ 0.6 & 0.027 & D03  &  4250 & 0.324 $\pm$ 0.000 & 19.5 $\pm$  0.7 \\
$\mu$ UMa      & 14.2 $\pm$ 0.5 & 0.009 & T09  &  3600 & 0.302 $\pm$ 0.001 & 10.5 $\pm$  0.7 \\
7 Cet          &  7.3 $\pm$ 0.3 & 0.045 & D03  &  3400 & 0.278 $\pm$ 0.001 &  8.3 $\pm$  0.1 \\
$\delta$ Oph   & 19.1 $\pm$ 0.2 & 0.098 & T09  &  3600 & 0.266 $\pm$ 0.003 & 12.2 $\pm$  0.2 \\
$\alpha$ Cet   & 13.1 $\pm$ 0.4 & 0.107 & T09  &  3500 & 0.320 $\pm$ 0.003 &  9.8 $\pm$  0.1 \\
$\beta$ Peg    & 16.6 $\pm$ 0.2 & 0.070 & D03  &  2800 & 0.213 $\pm$ 0.000 & 23.8 $\pm$  1.7 \\
\\
$\rho$ Per     & 10.6 $\pm$ 0.2 & 0.055 & D03  &  3300 & 0.228 $\pm$ 0.003 & 35.8 $\pm$  4.7 \\
$\pi$ Aur      &  4.3 $\pm$ 0.6 & 0.125 & T09  &  3800 & 0.348 $\pm$ 0.003 & 21.7 $\pm$  0.8 \\
$\delta$ Vir   & 16.4 $\pm$ 0.2 & 0.027 & T09  &  3100 & 0.296 $\pm$ 0.001 & 11.8 $\pm$  0.6 \\
$\beta$ Gru    & 18.4 $\pm$ 0.4 & 0.009 & T09  &  2950 & 0.293 $\pm$ 0.001 & 14.5 $\pm$  1.1 \\
\\
$\gamma$ Cru   & 36.8 $\pm$ 0.2 & 0.045 & T09  &  2850 & 0.308 $\pm$ 0.000 & 14.1 $\pm$  1.1 \\
$\delta^2$ Lyr &  4.4 $\pm$ 0.2 & 0.098 & D03  &  2200 & 0.424 $\pm$ 0.002 & 20.9 $\pm$  5.1 \\
57 Peg         &  4.2 $\pm$ 0.3 & 0.089 & T09  &  4350 & 0.272 $\pm$ 0.001 & 13.4 $\pm$  2.3 \\
TU CVn         &  4.7 $\pm$ 0.3 & 0.047 & D03  &  2150 & 0.393 $\pm$ 0.003 & 40.9 $\pm$  4.2 \\
2 Cen          & 17.8 $\pm$ 0.2 & 0.114 & D03  &  2250 & 0.391 $\pm$ 0.001 & 21.3 $\pm$  0.6 \\
R Lyr          & 10.9 $\pm$ 0.1 & 0.036 & T09  &  2100 & 0.361 $\pm$ 0.002 & 34.2 $\pm$  1.4 \\
\\
$\rho^1$ Ari   &  9.3 $\pm$ 0.3 & 0.304 & T09  &  2400 & 0.342 $\pm$ 0.006 & 25.2 $\pm$  3.2 \\
V537 Car       &  3.0 $\pm$ 0.5 & 0.214 & T09  &  1300 & 0.369 $\pm$ 0.002 & 45.7 $\pm$  1.8 \\
OP Her         &  3.4 $\pm$ 0.3 & 0.129 & D03  &  1450 & 0.238 $\pm$ 0.001 & 14.0 $\pm$  0.5 \\
NU Pav         &  6.9 $\pm$ 0.3 & 0.073 & D03  &  1800 & 0.345 $\pm$ 0.002 & 48.0 $\pm$  2.8 \\
\enddata
\tablenotetext{a}{References for extinction:  D03
  \citep{dri03}, T09 \citep{tab09}.}
\end{deluxetable*}

We then integrated the SiO absorption from 7.5~\mum\ to
10.7~\mum, accounting for the interstellar extinction 
profile based on our estimated $A_V$.  For the K giant 
sample, Paper~I measured the SiO band strength from 
7.3~\mum, but in the M giants, H$_2$O absorption at 
7.3~\mum\ forced us to shift to 7.5~\mum.
Figure~\ref{f.exsio} illustrates the technique for a 
sample spectrum.  

To estimate the uncertainties arising from our estimates of 
$A_V$, we measured the SiO and H$_2$O equivalent widths with 
$A_V$ set to zero.  We replaced the uncertainty in equivalent 
width with the difference between the two measurements if the 
difference was larger.  Tables~\ref{t.irsew} and 
\ref{t.swsew} give the results for the IRS and SWS samples, 
respectively.  Because of the effect of pointing on the
overall shape of the spectrum (Section~\ref{s.irs}), the
fitted temperatures in Table~\ref{t.irsew} are not physically
meaningful.  They are better interpreted as a guide to the
photometric accuracy of the spectrum in question.  We capped
the temperature at 10$^4$ K for HD~107893 and HD~177643,
which appear to be the most strongly affected by pointing.
The fitted temperatures in Table~\ref{t.swsew} are more
reasonable, if still of limited accuracy, due primarily to
the method used to conform the SWS spectra to the photometry
by \cite{eng06}.  The low apparent temperatures of the last 
seven M giants in Table~\ref{t.swsew} may be indicating the 
presence of dust.

Paper~I found that the $B-V$ color tracked the 
equivalent width of the SiO band better than spectral class 
for the K giants.  However, the behavior of $B-V$ versus 
spectral class is not monotonic for the M giants, as 
explained in Appendix~1.  Consequently, we will focus 
primarily on the behavior of the SiO, H$_2$O, and OH bands 
as a function of spectral class in this paper.

\subsection{Silicon monoxide} \label{s.sio} 

\begin{figure} 
\includegraphics[width=3.4in]{\figpath 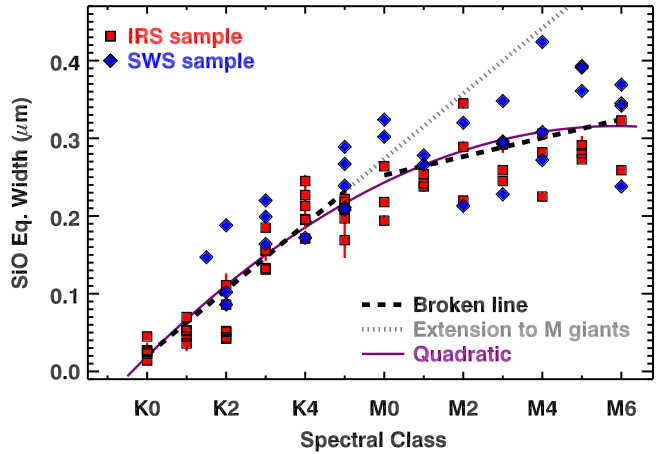} 
\caption{Equivalent width of the SiO band as a function of
spectral class.  The strength of the SiO band increases 
rapidly from K0 to K5, then more slowly from M0 to M6, with
considerable scatter at each spectral class.  In most cases,
the uncertainties are smaller than the plotting 
symbols.  The figure includes lines fitted separately to the 
K giants and M giants (dashed line), with the K-giant line
extended to later spectral classes (dotted line).  A quadratic
function (solid purple line) also follows the observed data
about as well as the broken linear fit.\label{f.spsio}}
\end{figure}

Figure~\ref{f.spsio} shows how the strength of the 
fundamental SiO band at 8~\mum\ depends on spectral class
for the full IRS sample of K and M giants and the
SWS sample.  Paper~I fit a quadratic to the relation for
just the IRS-observed K giants and the result was close
to linear.  Here, we fit a line to the K giants (dashed
line from K0 to K5).  Extending that line to later
spectral classes (dotted line) clearly overpredicts the
actual observed SiO equivalent widths for the M giants.

The slope of the line for the K giants is 0.042~\mum\ per
spectral class, while if we fit a separate line to spectral
classes M0--M6, the slope is only 0.012~\mum\ per class.
Concentrating on just the IRS sample, the standard deviation 
in equivalent width increases from $\sim$0.02~\mum\ for the 
K giants to over $\sim$0.03~\mum\ in the M giants.  Thus the 
change in SiO strength from one spectral class to the next is 
larger than the scatter in the K giants, but smaller in the 
M giants.

There is nothing unique about our choice to fit a broken
line to the data, with the break between K5 and M0.  As
long as the break is between K2/K3 and M0/M1, the $\chi^2$
residuals are about the same.  A quadratic (shown as a
purple curve in Figure~\ref{f.spsio}) also works about as
well.  Whatever the nature of the specific relationship,
the IRS spectra show that as the stars grow cooler, the
equivalent width of the SiO absorption band at 8~\mum\ 
increases more slowly with spectral class.  In addition,
the spread in SiO band strength is considerable in most 
spectral classes (from K2 on), and for the coolest giants in 
our sample, the spread is considerably larger than the 
increase in band strength from one subclass to the next.

\begin{figure} 
\includegraphics[width=3.4in]{\figpath 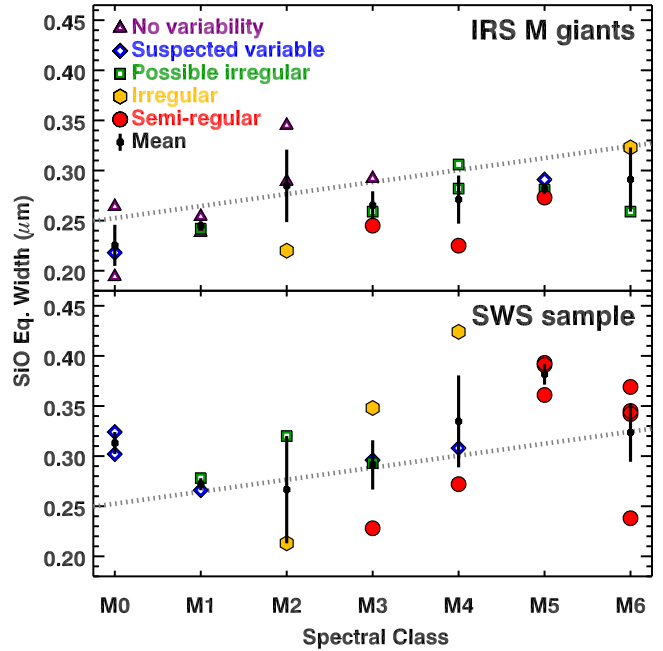} 
\caption{Equivalent width of the SiO band for the M giants,
color-coded by variability.  The dotted line is fitted to all 
of the plotted data and is the same in both panels.  The error 
bars for the mean equivalent widths are the standard deviations 
at each spectral class.  No dependence on variability is 
apparent.\label{f.varsio}}
\end{figure}

Figure~\ref{f.varsio} concentrates on just the M giants and
color codes them by their variability class.  The IRS and SWS
spectra show no evidence that more variable stars are 
associated with deeper SiO absorption bands.  The dotted line
in both panels is the same line fitted to the 40 M giants
observed by the IRS and SWS in Figure~\ref{f.spsio}.  To
compare the SiO band depth versus variability, we have 
grouped our sample into the more variable stars, which 
includes the 17 stars classified as irregulars or 
semi-regulars.  The other 23 stars classified as possible
irregulars, suspected variables, or with no identified
variability are the control sample.  For each group,
we measure the mean difference in actual equivalent width
versus the equivalent width expected from the fitted line
for that spectral class.  The control sample has a mean
difference of $-$1.0 nm, compared to an uncertainty in the
mean of 7.4 nm.  The more variable stars have a mean
difference of 0.8 $\pm$ 15.4 nm.  Thus we can conclude that
variability has virtually no impact on SiO equivalent width.

\subsection{Water vapor} \label{s.h2o} 

Most previous studies of emission or absorption from water 
vapor in the spectra of evolved stars have been of 
discrete lines, either in the vicinity of 6.5--6.7~\mum\ 
\citep[e.g.][]{tsu97,tsu01} or in the 12~\mum\ region 
\citep[e.g.][]{ryd02,ryd06}.  However, some of the spectra
in our sample of M giants show decidedly more absorption,
most notably BD+44 2199 in Figure~\ref{f.spirs2}, where the
absorption is strong enough from 6.4 to 7.5~\mum\ to leave 
what looks like an apparent emission feature at 
6.2--6.3~\mum. 

\begin{figure} 
\includegraphics[width=3.4in]{\figpath 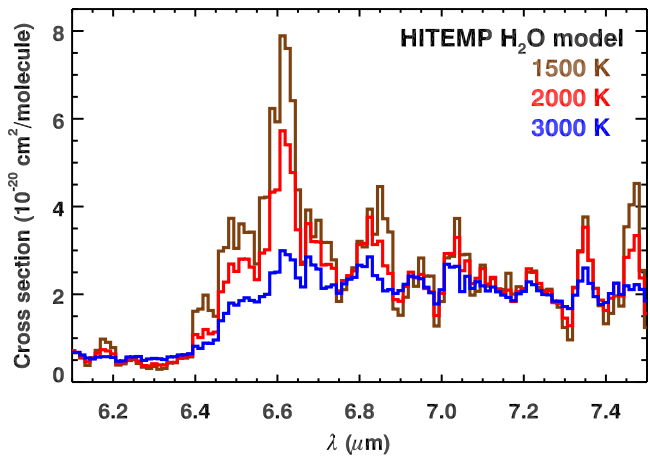} 
\caption{The absorption cross section of H$_2$O at three
temperatures.  These synthetic spectra are based on the 
{\rm Kspectrum} code and the HITEMP database \citep{rot10}, 
and they have been downsampled to a resolution about twice 
that of the IRS at these wavelengths.  As the temperature 
drops, the bands at $\sim$6.6 and 6.8~\mum\ grow more 
prominent.\label{f.modelh2o}}
\end{figure}

\begin{figure} 
\includegraphics[width=3.4in]{\figpath 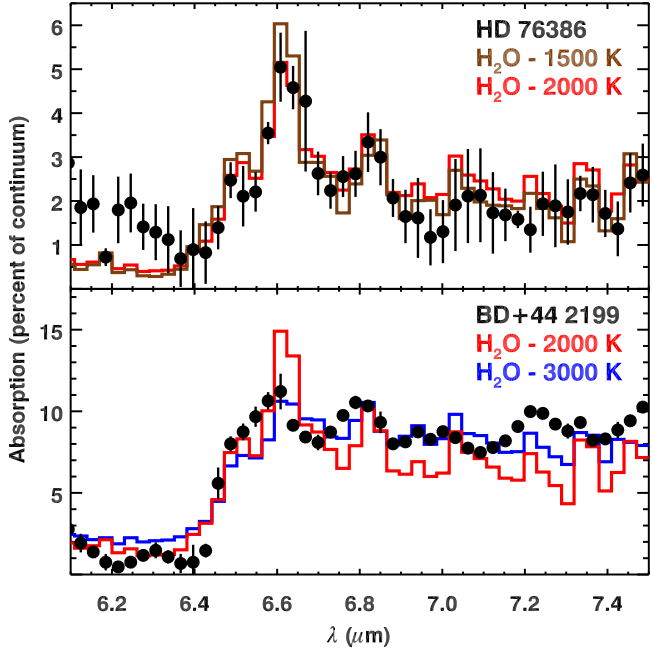} 
\caption{Spectra of the absorption in two spectra in our 
sample compared to the synthetic water vapor absorption in 
Fig.~\ref{f.modelh2o}.  {\it Top:} BD+44 2199 has the 
strongest H$_2$O absorption in our sample, and the minimal 
substructure at $\sim$6.6 and 6.8~\mum\ is most consistent 
with our synthetic spectrum at 3000 K, which is roughly the 
dissociation temperature of H$_2$O.  {\it Bottom:}  HD~76386 
is an example of the several spectra which show two apparent 
absorption bands from H$_2$O at $\sim$6.6 and 6.8~\mum, and 
these are generally more consistent with cooler temperatures.
\label{f.spech2o}}
\end{figure}

Figure~\ref{f.modelh2o} presents synthetic spectra of H$_2$O
generated with the line lists from the HITEMP database 
\citep{rot10} and the {\rm Kspectrum} software package 
\citep{wor10}.  HITEMP includes over 100 million water vapor
lines and is more suitable for warm stellar atmospheres than
databases such as HITRAN, which is designed for the Earth's
atmosphere \citep{rot12}.  {\rm Kspectrum} generates 
high-resolution spectra from line-by-line databases using 
Voigt and Lorentz line profiles without applying radiative
transfer methods.  It has been used primarily to generate 
synthetic spectra from planetary atmospheres 
\citep[e.g.][]{ram14}, and we use it to generate cross 
sections as a function of wavelength and temperature.  We 
found that pressure had little effect.  As the temperature of 
water vapor drops from 3000~K (roughly its dissociation 
temperature) to 2000~K, the absorption structure in the 
6--8~\mum\ region develops two prominent bands at $\sim$6.6 
and 6.8~\mum.

Figure~\ref{f.spech2o} compares the absorption structure
in two of our spectra to the synthetic H$_2$O spectra and
demonstrates that the structure in the IRS spectra at 
6--7~\mum\ is indeed from water vapor.  BD+44 2199 shows
the strongest H$_2$O absorption in our sample, and resembles
most closely the synthetic spectrum at 3000 K, because it
doesn't show a prominent 6.6~\mum\ feature.  The deep 
absorption in the observed spectrum requires a higher column 
density than the other spectra.  Many of the spectra with 
weaker water vapor absorption show two distinct bands at 
$\sim$6.6 and 6.8~\mum, which the cooler synthetic spectra 
fit well.  HD~76386 is a typical example, and the 2000~K 
model fits its spectrum better than the 1500 K model.  

The synthetic spectra firmly identify water vapor as the
carrier of the excess absorption between 6.4 and 7.5~\mum\ in
the M giants.  The spectra also show that varying the 
temperature of the water vapor can explain the different 
structures apparent in the 6--7~\mum\ range in our sample.  
These preliminary modeling results point to the potential for 
more detailed analysis, but that is beyond the scope of the 
present paper.

Figure~\ref{f.modelh2o} shows that the absorption from
H$_2$O does not go to zero in the 6.2--6.3~\mum\ region.
Nor does it go to zero at 7.5~\mum\ at the edge of the SiO
band.  Thus our measurement from 6.4--7.5~\mum\ is only a
partial measure of the H$_2$O absorption and should be
treated somewhat qualitatively.  

The water vapor also affects our continuum fitting, but
its impact on the measured SiO equivalent width is small.
Our fitting of an Engelke function to the continuum will
still effectively isolate the SiO band even if water vapor 
absorption shifts the entire continuum downward.
 and should still lead to a reliable measurement.

\begin{figure} 
\includegraphics[width=3.4in]{\figpath 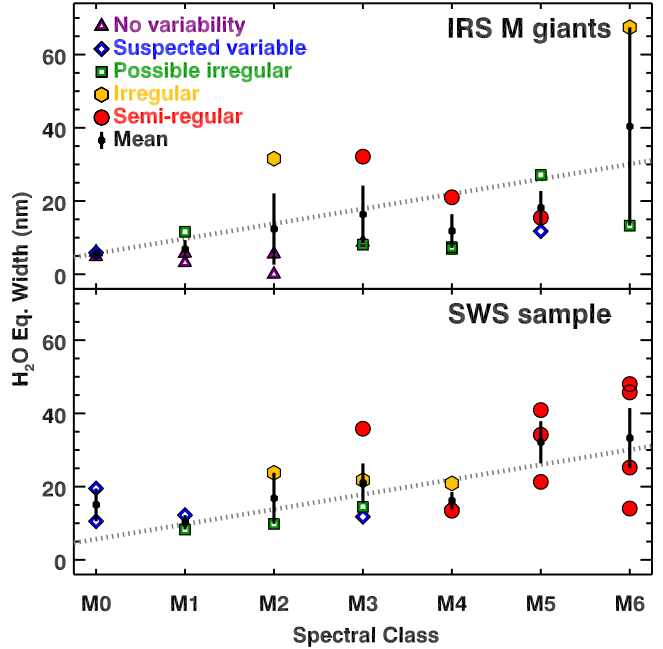} 
\caption{H$_2$O absorption in the M giants, color-coded by 
variability.  As in Fig.~\ref{f.varsio}, the dotted line is 
the same in both panels, and the error bars for the means at
each spectral class depict the standard deviations.  The 
more variable stars tend to have the stronger H$_2$O 
bands.\label{f.varh2o}}
\end{figure}

A comparison of Figures~\ref{f.spsws1} and \ref{f.spsws2}
show that the H$_2$O absorption is not visible in the SWS
targets of spectral class K.  Thus we only consider the 
equivalent widths for H$_2$O measured for the M 
giants.\footnote{For HD~177643, we treat the H$_2$O 
equivalent width as zero here and in the following analysis, 
because the negative value results from the distortion in
the spectrum due to mispointing and not {\it emission}.}
Figure~\ref{f.varh2o} plots the equivalent width as a
function of spectral class for the M giants in both the 
IRS and SWS samples.  The dotted line in both panels is
fitted to all of the data, and it shows a gentle rise, 
4.1 nm per spectral class, compared to a median standard
deviation in each class of 4.6 nm for the IRS sample and
5.4 nm for the SWS sample.  While the spread is substantial
compared to the slope, the net change in absorption strength
from M0 to M6 is 25 nm, or roughly five times the typical
scatter.

Figure~\ref{f.varh2o} shows that variable stars tend to have
stronger H$_2$O bands.  The IRS sample includes a range of
variabilities from M2 to M6, and in four of the five cases,
the more variable stars have stronger H$_2$O contributions.
In the SWS sample, only M2--M4 show a distribution of
variabilities, and in two of the three cases, the same
result holds.  

To be more rigorous, we can apply the same test used for the 
effect of variability on the SiO band and compare the mean 
difference from the fitted line for the 17 irregulars and 
semi-regulars to the 23 other less variable stars.  The first 
group has a mean difference of $+$6.5 $\pm$ 3.3 nm, while the 
second is $-$5.4 $\pm$ 1.8 nm.  Thus the more variable stars 
have equivalent widths typically 11.9 nm stronger than the 
less variable stars.  The combined uncertainty in this 
difference is 3.8 nm, making the difference between the two 
variability groups a 3.2-$\sigma$ result.

\subsection{Hydroxyl} \label{s.oh} 

The OH bands are faint, typically only 1--2\% of the 
continuum and requiring signal/noise ratios (SNRs)
well over 100 for a reasonable detection.  Consequently 
Paper~I limited their analysis of the OH bands in the
K giants to the five standards observed repeatedly over the
cryogenic {\it Spitzer} mission and five signifiantly 
brighter K giants observed for cross-calibration purposes.  
These ten stars revealed a strong dependence of OH band 
strength with spectral class in the K giants, as measured in 
the four bands between 14 and 17~\mum. 

\begin{figure} 
\includegraphics[width=3.4in]{\figpath 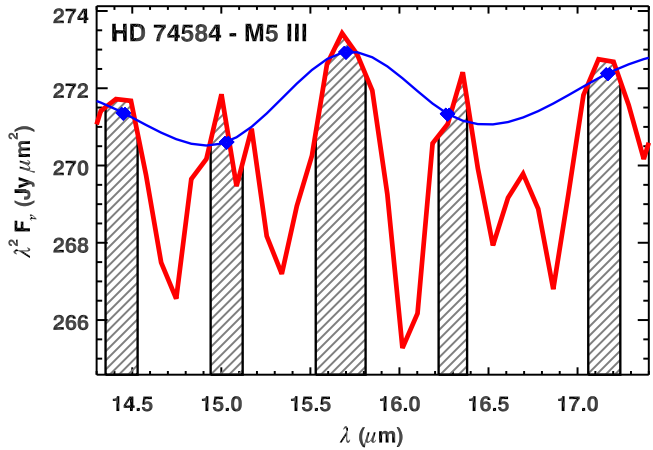} 
\caption{Extracting the OH absorption bands between 14.5 and 
17.1~\mum.  The continuum (blue) is estimated with a spline, 
using the data in the shaded regions.  The data between the 
shaded regions are the OH absorption bands.  The strongest 
OH band in this spectrum (at 16~\mum), is only $\sim$2\% of 
the continuum.\label{f.oh}}
\end{figure}

\begin{deluxetable}{lr} 
\tablenum{5}
\tablecolumns{2}
\tablewidth{0pt}
\tablecaption{OH equivalent widths in the IRS M giants}
\label{t.irsoh}
\tablehead{
  \colhead{Target} & \colhead{Eq.\ width} \\
  \colhead{ } & \colhead{14--17~\mum\ (nm)} 
}
\startdata
HD 13570       & 13.5 $\pm$ 3.0 \\
HD 19554       & 18.6 $\pm$ 1.9 \\
HD 107893      & 16.1 $\pm$ 3.9 \\
\\
HD 17678       & 21.6 $\pm$ 4.0 \\
BD+47 2949     & 13.2 $\pm$ 2.3 \\
HD 206503      & 21.1 $\pm$ 3.6 \\
\\
HD 122755      &  9.1 $\pm$ 3.0 \\
HD 177643      & 18.2 $\pm$ 2.3 \\
HD 189246      & 15.0 $\pm$ 3.0 \\
\\
HD 26231       & 21.4 $\pm$ 2.3 \\
HD 127693      & 20.3 $\pm$ 2.5 \\
HD 223306      &  9.5 $\pm$ 2.5 \\
\\
HD 17766       & 11.0 $\pm$ 2.8 \\
HD 32832       & 12.8 $\pm$ 2.4 \\
HD 46396       & 17.5 $\pm$ 2.3 \\
\\
HD 68422       & 18.1 $\pm$ 2.2 \\
HD 74584       & 20.3 $\pm$ 2.3 \\
HD 76386       & 19.1 $\pm$ 3.7 \\
\\
HD 8680        & 17.9 $\pm$ 3.5 \\
BD+44 2199     & 22.6 $\pm$ 5.2 \\
\enddata
\end{deluxetable}

Our single pointings at the 20 M giants had SNRs just 
sufficient enough to detect the OH bands.  We used the same 
approach as Paper~I to maintain consistency, determining the 
continuum in the intervals between the bands and then fitting 
a spline through the estimated continua.  Figure~\ref{f.oh} 
illustrates the process.  The example chosen, HD 74584, is 
one of the better-behaved spectra in this spectral region.
We tried several alternative approaches, such as coadding 
the spectra at each spectral class or forcing the spline 
continuum points to align with a Rayleigh-Jeans tail, with 
no improvement in the results.  Table~\ref{t.irsoh} presents 
the measured equivalent widths, summed from the four bands in 
the 14--17~\mum\ region.

\begin{figure} 
\includegraphics[width=3.4in]{\figpath 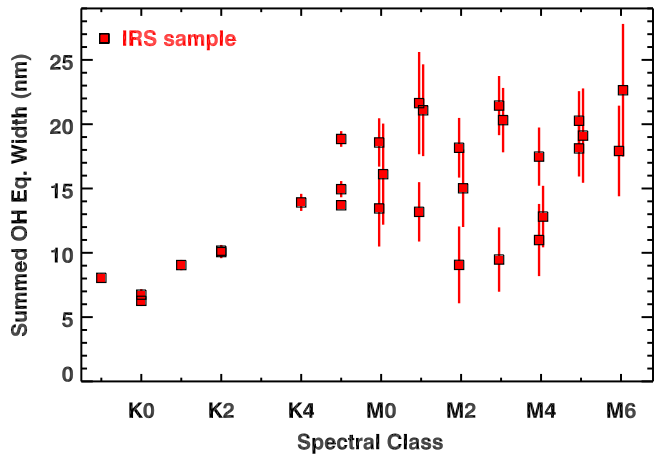} 
\caption{OH band strength summed from 14.5 to 17.1~\mum\ and
plotted as a function of spectral class for the sources 
observed with the IRS.  The OH bands grow stronger from K0 
to K5 and then plateau to later spectral classes.  The M
giants have been slightly offset horizontally to limit the
overlap of the error bars.\label{f.spoh}}
\end{figure}

\begin{figure} 
\includegraphics[width=3.4in]{\figpath 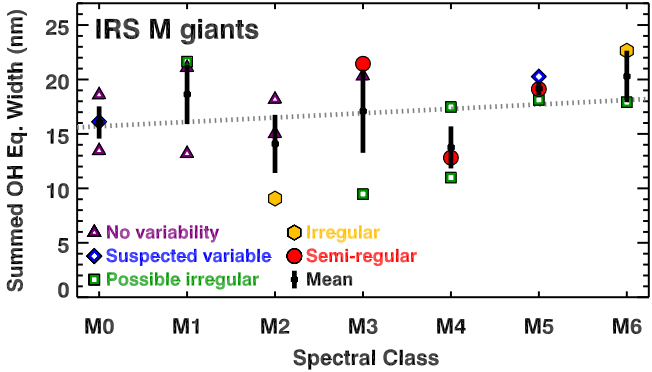} 
\caption{Equivalent width of the OH bands for the M giants, 
color-coded by variability.  The dotted line is fitted to 
just the data plotted in this panel and shows the lack of a 
strong dependence on spectral class in the M giants.  
Variability does not appear to be a factor, either.  The
error bars for the mean equivalent widths give the standard
deviations.\label{f.varoh}}
\end{figure}

Figure~\ref{f.spoh} shows the result of our extractions:
considerable scatter in each spectral class from M0 to M6,
with a median standard deviation of 2.4~nm.  The line fitted
to the M giants observed by the IRS has a slope of only 
0.4~nm per spectral class, making the apparent change in
band strength from M0 to M6 comparable to the typical scatter
in each spectral class.  The IRS spectra do not reveal any
significant dependence of equivalent width on spectral class
for the M giants.  Figure~\ref{f.varoh} shows that 
variability is not responsible for the scatter in OH band 
strength.  For both the five variables and the 15 stars in 
the control sample, the mean difference between actual 
equivalent widths and those expected from the fitted line
are statistically insignificant ($\sim$0.1~$\sigma$).

\section{Discussion} \label{s.disc} 

\subsection{SiO and OH absorption} 

Section~\ref{s.sio} showed that the equivalent width of
the fundamental SiO band at 8~\mum\ generally increased
toward later spectral classes, but with considerable scatter
and a decrease in the slope with later spectral class.
\cite{her02} previously noted each of these points, and thus 
our results fully confirm theirs.  Because we have more 
carefully constrained the IRS sample by luminosity class, we 
can conclude that the scatter is not due to luminosity.  Nor 
is it due to variability.  Paper~I suggested that 
metallicity could possibly influence the equivalent width of 
the SiO band and produce the observed scatter in a given
spectral class.  While that hypothesis is still plausible,
for both K and M giants, it remains untested.

We have detected OH bands in all of the M giants observed 
with the IRS, and while the OH band strength increases with 
later spectral types in the K giants, the dependence is 
much flatter in the M giants and shows substantial 
scatter.  We have detected no effect of stellar variability
on OH band strength.  While we were able to detect OH 
absorption bands in all 20 of the M giants observed by the
IRS, the spectra are right at the threshold for useful
analysis.  Higher quality data, with better resolutions and
SNRs, would help substantially in our understanding of how
the OH bands behave and depend on stellar properties.

\subsection{H$_2$O absorption} 

Previous identifications of H$_2$O in the 6--7~\mum\ region
were based on individual lines in higher-resolution data
\citep[e.g.][]{tsu97,tsu01}.  Our synthetic spectra show that 
H$_2$O is responsible for the broad absorption structure 
apparent in most of our M-giant spectra from 6.3~\mum\ to the
SiO bandhead at 7.5~\mum.

The synthetic spectra demonstrate that the HITEMP line list 
is complete enough to support efforts to model the spectra in
this wavelength regime, although some discrepancies between 
the observations and the synthetic spectra are still 
apparent, especially to the red of $\sim$7~\mum.  For 
example, both panels of Figure~\ref{f.spech2o} show an 
absorption feature at 7.1~\mum\ in the synthetic spectra 
which is not apparent in the observed data.  

Our limited modeling effort suggests that the different
spectral structures seen in the 6.4--6.8~\mum\ region can
be explained by different temperatures.  In the example 
presented in Figure~\ref{f.spech2o}, BD+44 2199 requires a 
higher column density and warmer temperatures compared to 
HD~76386.  Figure~\ref{f.modelh2o} shows that the relative 
strengths of the 6.6 and 6.8~\mum\ bands might serve as a 
temperature diagnostic.  In addition, the position of the 
6.8~\mum\ band appears to shift to the red with cooler 
temperatures.  Further modeling is required to 
substantiate these tentative conclusions.

Stars which are more variable tend to have stronger absorption 
from H$_2$O in the 6.3--7.5~\mum\ range.  Thus stellar 
variability plays a role in the formation of H$_2$O, but not
OH or SiO.

The IRS and SWS data show a gentle but measurable increase in 
H$_2$O band strength with spectral class in the M giants.  
This result differs from the previous conclusion of 
\cite{ard10}, who found no apparent change in band strength 
with spectral class, but it should be kept in mind that they 
examined a smaller number of M giants than the current 
sample, and the stars in their sample tended to be relatively
non-variable.

\cite{tsu00} detected water vapor emission in the spectrum
of the supergiant $\mu$~Cep and proposed that it arose from 
an extended molecular sphere, which he identified as a 
MOLsphere.  \cite{tsu01} added several red giants to the list 
of possible MOLsphere sources, but the idea remains 
controversial.  High-resolution spectra of $\alpha$~Boo in 
the 11--12~\mum\ range show absorption lines with 
temperatures more consistent with the photosphere of the 
star than a detached molecular layer \citep{ryd02}.  This 
star is much warmer than the typical H$_2$O absorber in our 
samples, but \cite{ryd06} found a similar result at 
$\sim$12~\mum\ in $\mu$~Cep, which has a spectral type of 
M1--2 Ia--Iab.  Further high-resolution spectra show similar 
results for a larger sample of K and M giants \citep{ryd15},
with generally stronger absorption than expected and
temperatures consistent with a cool photosphere.

For geometric reasons, water vapor in an extended molecular
sphere should produce emission lines, or at a minimum
substantially weakened absorption lines.  \cite{tsu09}
modified his original MOLsphere hypothesis to account for 
this lack of emission lines by suggesting that the molecules 
may exist in clouds in the outer atmospheres of the stars.  
In our sample of 20 M giants observed by the IRS and 20 
observed by the SWS, not one shows water vapor in emission.  
That result places some limits on the patchiness of the 
clouds and how far above the photosphere they could lie.

\subsection{Considering the photometry} 

\begin{figure} 
\includegraphics[width=3.4in]{\figpath 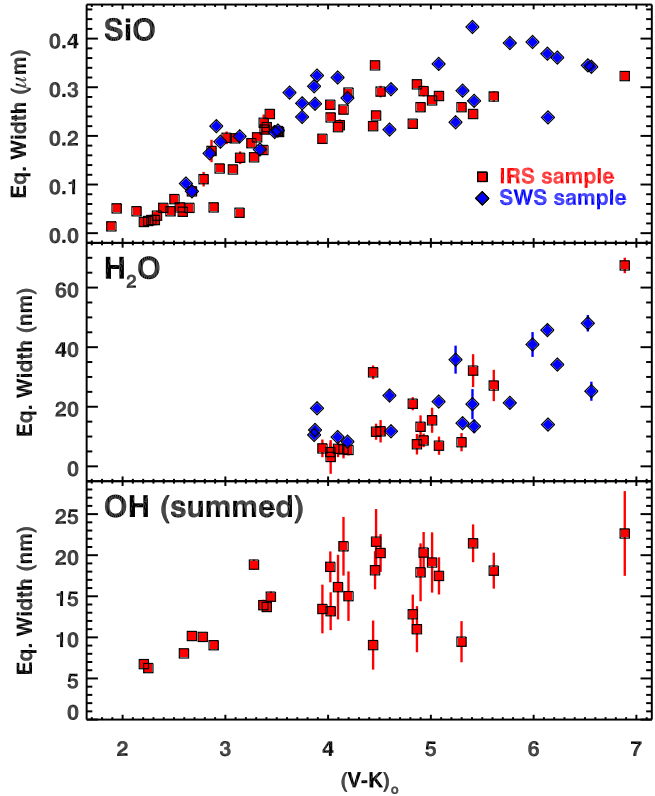} 
\caption{The equivalent width of the SiO band, measured
H$_2$O, and summed OH bands as a function of dereddened
$V-K$ color for the IRS and SWS samples.\label{f.col3}}
\end{figure}

Figure~\ref{f.col3} plots the equivalent widths of the
infrared molecular bands versus $(V-K)_0$.  Appendix~1
explains how we determined the photometric colors for our
sample and shows that $B-V$, which Paper~I found to be 
a useful diagnostic for K giants, is double-valued in M 
giants.  The $V-K$ color reddens monotonically with spectral 
class and can give us more insight on the behavior of the 
molecular bands.  

For SiO as a function of $V-K$ color, the change in slope 
between the K and M giants is apparent, just as when plotting 
versus spectral class.  In Figure~\ref{f.spsio}, a fair 
fraction of the SWS sources were above the line segments 
fitted to the combined sample, but in the top panel of 
Figure~\ref{f.col3}, these sources have moved closer to the 
rest of the population.  At the red end of the sample, six 
out of the seven SWS sources at M4--M6 with strong SiO 
absorption (equivalent width $>$ 0.34~\mum), have the reddest 
$V-K$ colors in the sample.  Only BD+44 2199 in the IRS 
sample is redder.  Plotting versus color instead of spectral 
class aligns these sources better with the trends shown by 
the rest of the sample.

\subsection{Towards dust production} 

The SWS sample represents a sizable subset of the red giants
closest to the Sun, and it shows how the percentage of
variables increases quickly past a spectral class of M2.
All of the SWS sources of spectral class M4--M6 are 
irregulars or semi-regulars.  When selecting the IRS sample, 
we tried to avoid known variables, but we were unable to keep 
them out of the sample altogether.  Irregular and 
semi-regular variables are long-period variables typically 
associated with the thermally pulsing asymptotic giant branch 
(AGB), and their presence in our sample reveals that we are 
looking at two groups of stars.  AGB stars appear to dominate 
spectral classes past M4.  Earlier spectral classes could be 
either making their first ascent of the red-giant branch 
(RGB) or are early AGB stars.

The likely presence of dust in several of the reddest
sources in our sample also suggests that it includes AGB 
stars beginning the mass-loss process.  As previously noted, 
several of the reddest sources show inflections in their 
spectra suggestive of alumina dust.  BD+44 2199 has the 
reddest $V-K$ color in the sample, the strongest H$_2$O
absorption, the second strongest SiO absorption, and its
spectrum shows an increasing excess with wavelength to the
red of 10~\mum.  Both its $V-K$ color and its mid-infrared
spectrum indicate that it is producing dust.
The reddest sources in the SWS sample account for most of 
the remaining strong SiO and H$_2$O absorbers.  These are
also likely to be thermally pulsing AGB stars beginning to
lose mass and produce dust.

Our spectra reveal the molecular precursors to the dust we
expect to form as these stars continue to evolve, and they
point to the potential for further observations.  Deeper 
integrations and higher spectral resolutions would reveal 
in more detail the physical conditions of the OH and H$_2$O 
molecules and better constrain their physical locations in
the stellar atmospheres.
In particular, the H$_2$O bands at these wavelengths look to 
have strong diagnostic possibilities.  New samples in systems
with known distances and metallicities may be the best means
of investigating the origin of the scatter in the present
sample.

\section{Summary} \label{s.conclude} 

We have observed 20 M giants with the IRS on {\it Spitzer}.
These spectra, combined with our previous IRS sample of
33 K giants and the SWS sample of 13 K-type stars and
20 M giants, reveal how the strength of the SiO, OH, and
H$_2$O bands depend on spectral class and $V-K$ color.

The equivalent width of the SiO band at 8~\mum\ increases
as the stars grow cooler, but it increases more gradually
at later spectral classes.  The scatter is considerable and
intrinsic to the sample.  These results confirm the earlier
study by \cite{her02}.  The scatter does not result from 
differences in the luminosities or variability properties of
the stars.

Our synthetic spectra confirm that the structure in our
spectra between 6.3 and 7.5~\mum\ is from H$_2$O absorption.
These bands are not easily detected in the K giants, but in
the M giants, they increase in strength as the stars grow
cooler, but again, with considerable scatter.  In this
case, the scatter arises from the variability properties of
the stars, with more variable sources generally showing
stronger absorption in a given spectral class.

The OH bands at 14--17~\mum\ climb in strength from K0 to K5,
but for cooler stars show little obvious dependence on
temperature.  The scatter in equivalent widths in the M 
giants is considerable.  Variability plays no obvious role on
OH band strength.

\acknowledgements

The newly reported observations were made with the {\it 
Spitzer Space Telescope}, which is operated by the Jet 
Propulsion Laboratory, California Institute of Technology, 
under NASA contract 1407.  NASA provided support of this work 
through contract 1257184 issued by JPL/Caltech.  We thank 
the science editor, D.\ Gies, and the anonymous referee, whose
input led to a much-improved paper, and G.\ van Belle, whose
comments helped us better utilize the photometry.  This 
research has made use of NASA's Astrophysics Data System and 
the SIMBAD database, which is operated at Centre de 
Donn\'{e}es astronomiques in Strasbourg, France.  The Infrared 
Science Archive (IRSA) at Caltech has also proven to be 
extremely helpful.

\section*{Appendix 1.  Photometry}

Paper~I found that the $B-V$ color of the K giants in 
their sample tracked the equivalent width of the SiO band at
8~\mum\ somewhat better than the spectral class.  They 
proposed modifying some of the spectroscopically assigned
spectral classifications on this basis.  It follows that we
should investigate the photometric properties of the SWS 
sample and the M giants observed by the IRS.

\begin{deluxetable*}{lrrlccllll} 
\tablenum{6}
\tablecolumns{10}
\tablewidth{0pt}
\tablecaption{Photometry for the SWS sample}
\label{t.swsphot}
\tablehead{
  \colhead{Target} & \colhead{$B$} & \colhead{$V$} & \colhead{Optical} & 
  \colhead{$B_T$} & \colhead{$V_T$} & \colhead{$J$} & \colhead{$H$} & 
  \colhead{$K$} & \colhead{Near-IR} \\
  \colhead{ } & \colhead{(mag)} & \colhead{(mag)} & 
  \colhead{Reference\tablenotemark{a}} &
  \colhead{(mag)} & \colhead{(mag)} & \colhead{(mag)} & \colhead{(mag)} &
  \colhead{(mag)} & \colhead{Reference\tablenotemark{b}}
}
\startdata
$\alpha$ Boo   &    1.18 & $-$0.05 & J66, L70 &       \nodata      & 0.161 $\pm$ 0.014 & $-$2.229 $\pm$ 0.100   & $-$2.938 $\pm$ 0.078   & $-$2.961 $\pm$ 0.212   & CIO \\
$\gamma^1$ And &    3.63 &    2.26 & L71      &  3.984 $\pm$ 0.014 & 2.308 $\pm$ 0.009 & $-$0.030 $\pm$ 0.079   & $-$0.715 $\pm$ 0.021   & $-$0.840 $\pm$ 0.048   & CIO \\
$\alpha$ Ari   &    3.15 &    2.00 & J66, L70 &  3.487 $\pm$ 0.014 & 2.125 $\pm$ 0.009 & ~~~0.098 $\pm$ 0.136   & $-$0.529 $\pm$ 0.036   & $-$0.634 $\pm$ 0.039   & CIO \\
$\xi$ Dra      &    4.93 &    3.75 & J66      &  5.253 $\pm$ 0.014 & 3.853 $\pm$ 0.009 & ~~~1.767 $\pm$ 0.022   & ~~~1.192 $\pm$ \nodata & ~~~1.038 $\pm$ 0.029   & CIO \\
\\
$\sigma$ Oph   &    5.83 &    4.33 & J66      &  6.293 $\pm$ 0.014 & 4.496 $\pm$ 0.009 & ~~~1.815 $\pm$ 0.092   & ~~~1.100 $\pm$ \nodata & ~~~1.003 $\pm$ 0.015   & CIO \\
$\lambda$ Gru  &    5.83 &    4.46 & J66      &  6.270 $\pm$ 0.014 & 4.618 $\pm$ 0.009 & ~~~2.264 $\pm$ 0.030   & ~~~1.436 $\pm$ 0.248   & ~~~1.506 $\pm$ 0.027   & P10, 2MASS \\
$\alpha$ Tuc   &    4.25 &    2.86 & J66      &  4.679 $\pm$ 0.014 & 2.998 $\pm$ 0.009 & ~~~0.558 $\pm$ \nodata & $-$0.080 $\pm$ \nodata & $-$0.090 $\pm$ \nodata & CIO \\
$\beta$ UMi    &    3.55 &    2.08 & J66      &  3.998 $\pm$ 0.014 & 2.215 $\pm$ 0.009 & $-$0.450 $\pm$ \nodata & ~~~~~~~~~~~~\nodata    & $-$1.315 $\pm$ 0.106   & CIO \\
\\
$\delta$ Psc   &    5.95 &    4.44 & J66      &  6.399 $\pm$ 0.015 & 4.594 $\pm$ 0.009 & ~~~1.777 $\pm$ 0.056   & ~~~1.000 $\pm$ 0.028   & ~~~0.857 $\pm$ 0.049   & CIO \\
$\gamma$ Phe   &    4.98 &    3.41 & J66      &  5.496 $\pm$ 0.014 & 3.613 $\pm$ 0.009 & ~~~0.542 $\pm$ 0.013   & ~~~~~~~~~~~~\nodata    & $-$0.360 $\pm$ 0.010   & P10 \\
$\alpha$ Tau   &    2.40 &    0.86 & L70      &  2.937 $\pm$ 0.006 & 1.160 $\pm$ 0.011 & $-$1.887 $\pm$ 0.067   & $-$2.628 $\pm$ 0.100   & $-$2.825 $\pm$ 0.198   & CIO \\
H Sco          &    5.73 &    4.16 & J66      &  6.238 $\pm$ 0.015 & 4.346 $\pm$ 0.009 & ~~~1.326 $\pm$ 0.021   & ~~~~~~~~~~~~\nodata    & ~~~0.394 $\pm$ 0.018   & P10 \\
$\gamma$ Dra   &    3.74 &    2.22 & J66      &  4.246 $\pm$ 0.014 & 2.381 $\pm$ 0.009 & $-$0.443 $\pm$ 0.043   & $-$1.160 $\pm$ \nodata & $-$1.340 $\pm$ 0.065   & CIO \\
\\
$\beta$ And    &    3.62 &    2.05 & J66, L70 &  4.155 $\pm$ 0.014 & 2.244 $\pm$ 0.009 & $-$0.869 $\pm$ 0.078   & $-$1.667 $\pm$ 0.067   & $-$1.873 $\pm$ 0.074   & CIO \\
$\mu$ UMa      &    4.64 &    3.05 & J66      &  5.126 $\pm$ 0.014 & 3.216 $\pm$ 0.009 & ~~~0.101 $\pm$ 0.012   & $-$0.685 $\pm$ 0.006   & $-$0.856 $\pm$ 0.043   & CIO \\
7 Cet          &    6.12 &    4.46 & J66      &  6.593 $\pm$ 0.015 & 4.629 $\pm$ 0.009 & ~~~1.298 $\pm$ \nodata & ~~~0.369 $\pm$ \nodata & ~~~0.191 $\pm$ \nodata & CIO \\
$\delta$ Oph   &    4.34 &    2.75 & J66      &  4.808 $\pm$ 0.014 & 2.896 $\pm$ 0.009 & $-$0.241 $\pm$ 0.052   & $-$1.030 $\pm$ 0.049   & $-$1.266 $\pm$ 0.058   & CIO \\
$\alpha$ Cet   &    4.17 &    2.53 & L70      &  4.687 $\pm$ 0.014 & 2.716 $\pm$ 0.009 & $-$0.629 $\pm$ 0.075   & $-$1.462 $\pm$ 0.156   & $-$1.683 $\pm$ 0.058   & CIO \\
$\beta$ Peg    &    4.15 &    2.50 & L70      &  4.610 $\pm$ 0.014 & 2.654 $\pm$ 0.009 & $-$1.120 $\pm$ 0.065   & $-$1.997 $\pm$ 0.066   & $-$2.214 $\pm$ 0.067   & CIO \\
\\
$\rho$ Per     &    5.04 &    3.39 & J66      &  5.372 $\pm$ 0.014 & 3.539 $\pm$ 0.009 & $-$0.792 $\pm$ 0.073   & $-$1.760 $\pm$ \nodata & $-$1.940 $\pm$ 0.035   & CIO \\
$\pi$ Aur      &    5.97 &    4.25 & J66      &  6.484 $\pm$ 0.014 & 4.516 $\pm$ 0.009 & ~~~0.268 $\pm$ 0.039   & $-$0.610 $\pm$ \nodata & $-$0.882 $\pm$ 0.054   & CIO \\
$\delta$ Vir   &    4.97 &    3.38 & J66, L70 &  5.382 $\pm$ 0.014 & 3.577 $\pm$ 0.009 & $-$0.203 $\pm$ 0.070   & $-$1.058 $\pm$ 0.049   & $-$1.244 $\pm$ 0.048   & CIO \\
$\beta$ Gru    &    3.73 &    2.11 & J66      &  4.131 $\pm$ 0.014 & 2.287 $\pm$ 0.009 & $-$2.126 $\pm$ 0.016   & $-$3.120 $\pm$ \nodata & $-$3.220 $\pm$ \nodata & P10, CIO \\
\\
$\gamma$ Cru   &    3.22 &    1.63 & J66      &       \nodata      &      \nodata      & $-$2.145 $\pm$ 0.205   & $-$2.880 $\pm$ 0.200   & $-$3.123 $\pm$ 0.088   & CIO \\
$\delta^2$ Lyr &    5.97 &    4.30 & J66      &  6.302 $\pm$ 0.014 & 4.464 $\pm$ 0.009 & $-$0.056 $\pm$ 0.043   & $-$0.913 $\pm$ 0.110   & $-$1.219 $\pm$ 0.042   & CIO \\
57 Peg         &    6.58 &    5.11 & J66      &  6.886 $\pm$ 0.015 & 5.294 $\pm$ 0.009 & ~~~0.785 $\pm$ 0.007   & $-$0.055 $\pm$ 0.092   & $-$0.360 $\pm$ 0.085   & CIO \\
TU CVn         &    7.39 &    5.84 & N78      &  7.730 $\pm$ 0.015 & 6.022 $\pm$ 0.009 & ~~~0.975 $\pm$ 0.021   & ~~~0.090 $\pm$ \nodata & $-$0.180 $\pm$ 0.062   & CIO \\
2 Cen          &    5.69 &    4.19 & J66      &  6.023 $\pm$ 0.014 & 4.416 $\pm$ 0.009 & $-$0.490 $\pm$ 0.021   & $-$1.398 $\pm$ 0.077   & $-$1.611 $\pm$ 0.138   & CIO \\
R Lyr          &    5.59 &    4.00 & J66      &  5.998 $\pm$ 0.014 & 4.355 $\pm$ 0.009 & $-$0.922 $\pm$ 0.047   & $-$1.803 $\pm$ 0.008   & $-$2.069 $\pm$ 0.045   & CIO \\
\\
$\rho^1$ Ari   &    7.44 &    5.93 & L70      &  7.416 $\pm$ 0.015 & 5.951 $\pm$ 0.010 & ~~~0.218 $\pm$ 0.074   & $-$0.735 $\pm$ 0.047   & $-$1.020 $\pm$ 0.027   & CIO \\
V537 Car       & \nodata & \nodata & \nodata  &  8.429 $\pm$ 0.016 & 6.847 $\pm$ 0.010 & ~~~1.410 $\pm$ 0.020   & ~~~0.280 $\pm$ 0.306   & ~~~0.368 $\pm$ 0.016   & P10, 2MASS \\
OP Her         &    7.93 &    6.32 & L70      &  8.211 $\pm$ 0.016 & 6.506 $\pm$ 0.010 & ~~~1.240 $\pm$ 0.042   & ~~~0.355 $\pm$ 0.024   & ~~~0.080 $\pm$ 0.054   & CIO \\
NU Pav         &    6.66 &    5.13 & H70, N78 &  6.815 $\pm$ 0.015 & 5.235 $\pm$ 0.009 & $-$0.250 $\pm$ 0.028   & $-$1.220 $\pm$ 0.071   & $-$1.510 $\pm$ 0.014   & CIO \\
\enddata
\tablenotetext{a}{Optical references:  H70 \citep{ho70}, 
  J66 \citep{joh66}, L70 \citep{lee70}, L71 \citep{lut71}, 
  N78 \citep{nic78}.}
\tablenotetext{b}{Near-infrared references:  2MASS 
  \citep[2MASS All-Sky Point-Source Catalog;][]{skr06},
  CIO \citep[Catalog of Infrared Observations, Ver.\ 
  5.1;][]{gez00}, P10 \citep{pri10}.}
\end{deluxetable*}

We used a similar procedure to that applied to the K
giants, starting with Tycho magnitudes \citep{hog00}, then 
converting to colors and magnitudes in the Johnson system.  
Paper~I used data from \cite{bes00} to determine the 
following conversion:
\begin{equation}
(B-V)_0 = 0.099 + 0.779 (B_T-V_T)_0,
\end{equation}
for 1.1 $\le$ $B_T-V_T$ $\le$ 1.9.  However, most of our
M giants are beyond the red limit.  We can use the SWS sample
to check the relation between Tycho and Johnson colors for
the M giants.  Table~\ref{t.swsphot} gives $B$ and $V$ 
magnitudes in both the Johnson and Tycho systems for the SWS 
sample.  These sources are bright enough that the Johnson 
magnitudes are from the seminal paper on the subject 
\citep{joh66}!  

Table~\ref{t.swsphot} also includes $JHK$ magnitudes, but 
here the brightness of the sources makes the task more 
difficult, because all of the sources are saturated in the 
2MASS survey.  To avoid the inaccuracies of the corrections 
for saturation, we have turned to the Catalog of Infrared 
Observations \citep[CIO, version 5.1;][]{gez00}.\footnote{The 
CIO, fondly known as the Galactic Phonebook, is available at 
http://ircatalog.gsfc.nasa.gov.}  When the CIO provides
multiple entries in a given filter, we determine the mean
and standard deviation after rejecting outliers from the 
sample.  We have not made a distinction between different
filter sets, for example just reporting $K$ magnitudes
intead of the usual 2MASS $K_s$.  When the CIO does not
provide photometry, we turned to the recalibration of bright
sources observed by the Diffuse Infrared Background
Experiment (DIRBE) on the {\it Cosmic Background Explorer}
({\it COBE}) by \cite{pri10}.  We have treated DIRBE bands
1 and 2 as roughly equivalent to $J$ and $K$.

\begin{figure} 
\includegraphics[width=3.4in]{\figpath 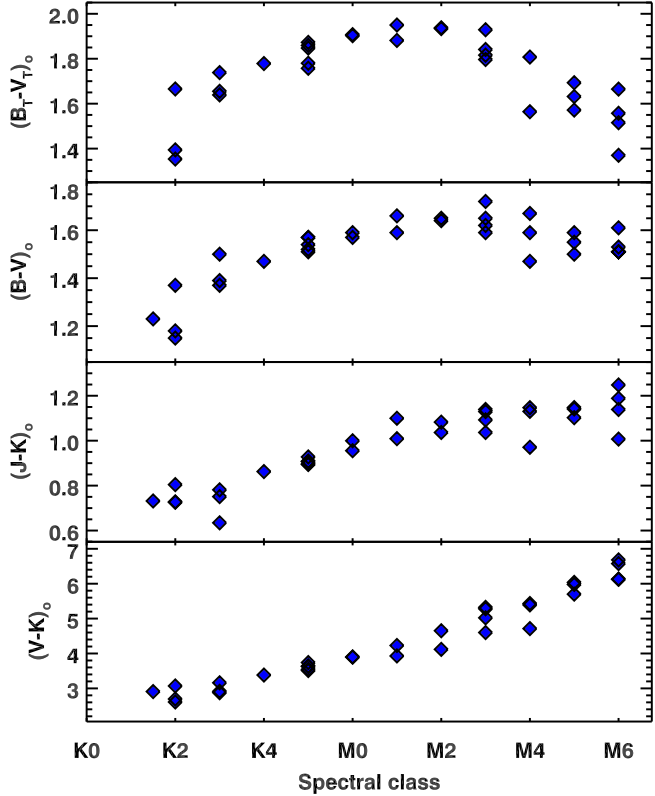} 
\caption{Optical and near-infrared colors as a function of 
spectral class for the SWS sample, for which we have 
reliable photometry for all of the colors 
plotted.\label{f.swsspcol}}
\end{figure}

Figure~\ref{f.swsspcol} plots $(B_T-V_T)_0$, $(B-V)_0$, 
$(J-K)_0$, and $(V-K)_0$ versus spectral class.  All of the
colors have been dereddened using the $A_V$ estimates in
Table~\ref{t.swsew} and the extinctions of \cite{rl85},
interpolating for the wavelengths of $B_T$ and $V_T$.  Both
$(B_T-V_T)_0$ and $(B-V)_0$ reach a maximum at $\sim$M2,
then decrease for later spectral classes.  After the
turnover, their slopes differ, with $(B_T-V_T)_0$ falling
more steeply than $(B-V)_0$.  

\begin{figure} 
\includegraphics[width=3.4in]{\figpath 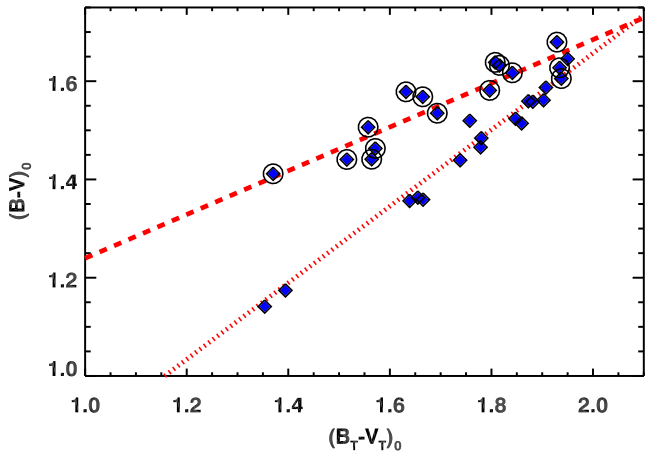} 
\caption{Converting from $(B_T-V_T)_0$ to $(B-V)_0$.  Two
conversions are required because later-type stars follow a 
different sequence.  The dotted line is from Paper~I and 
applies to the K giants.  The dashed line is fitted to all 
stars of spectral class M2 or later (which are circled).
\label{f.colconv}}
\end{figure}

As a result, the conversion from $(B_T-V_T)_0$ to $(B-V)_0$ 
is double-valued, as Figure~\ref{f.colconv} shows.  Spectral
classes up to and including M1 follow the relation from
Paper~I closely.  But M2 and later giants follow a
different relation:
\begin{equation}
(B-V)_0 = 0.795 + 0.445 (B_T-V_T)_0. 
\end{equation}
Because the $(B_T-V_T)_0$ color peaks at $\sim$1.9, a color
limit on whether to use Equations (1) or (2) is insufficient.
Either a break at spectral class M2 can be imposed, or
the $(V-K)_0$ color can be used.  M2 corresponds to $(V-K)_0$
$\sim$ 4.5.  (see Figure~\ref{f.swsspcol}, bottom panel).

\begin{figure} 
\includegraphics[width=3.4in]{\figpath 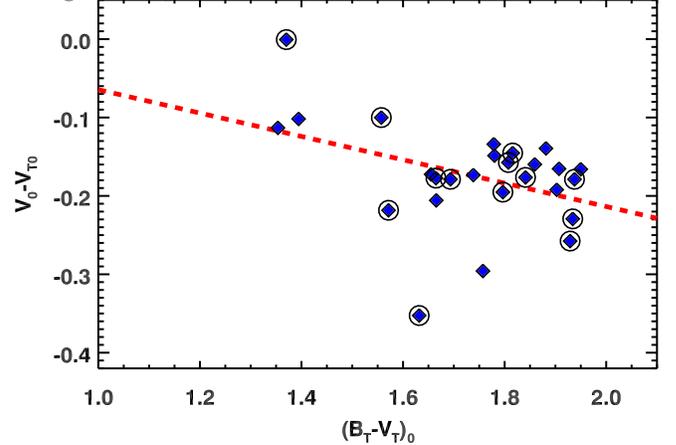} 
\caption{The conversion from $V_T$ to $V$ for late-type
stars, plotted as $V_0-V_{T0}$ versus $(B_T-V_T)_0$.
Data for M2 and later classes are circled.\label{f.vconv}}
\end{figure}

Figure~\ref{f.vconv} plots the conversion from $V_{T0}$ to
$V_0$ as a function of $(B_T-V_T)_0$:
\begin{equation}
V_0 - V_{T0} = 0.0845 - 0.149 (B_T-V_T)_0. 
\end{equation}
The noise in the data is considerable.  It is not related to 
the double-valued relation between $(B_T-V_T)_0$ and 
$(B-V)_0$, as the later spectral classes are not 
preferentially above or below the line fitted to all of the
data.

\begin{deluxetable*}{lllllll} 
\tablenum{7}
\tablecolumns{10}
\tablewidth{0pt}
\tablecaption{Photometry for the bright IRS sample}
\label{t.brightphot}
\tablehead{
  \colhead{Target} & \colhead{$B_T$} & \colhead{$V_T$} & 
  \colhead{$J$} & \colhead{$H$} & \colhead{$K$} & \colhead{Near-IR} \\
  \colhead{ } & \colhead{(mag)} & \colhead{(mag)} & 
  \colhead{(mag)} & \colhead{(mag)} & \colhead{(mag)} & 
  \colhead{Reference\tablenotemark{a}}
}
\startdata
$\delta$ Dra   &  4.322 $\pm$ 0.014 & 3.165 $\pm$ 0.009 & ~~~1.191 $\pm$ 0.020 & ~~~0.837 $\pm$ 0.166   & ~~~0.464 $\pm$ 0.026 & P10, 2MASS \\
42 Dra         &  6.345 $\pm$ 0.014 & 4.955 $\pm$ 0.009 & ~~~2.696 $\pm$ 0.029 & ~~~2.197 $\pm$ 0.196   & ~~~2.019 $\pm$ 0.043 & P10, 2MASS \\
$\xi$ Dra      &  5.253 $\pm$ 0.014 & 3.853 $\pm$ 0.009 & ~~~1.767 $\pm$ 0.022 & ~~~1.192 $\pm$ \nodata & ~~~1.038 $\pm$ 0.029 & CIO \\
HR 5755        &  7.809 $\pm$ 0.015 & 6.078 $\pm$ 0.009 & ~~~3.262 $\pm$ 0.138 & ~~~1.855 $\pm$ 0.212   & ~~~2.432 $\pm$ 0.126 & P10, 2MASS \\
$\gamma$ Dra   &  4.246 $\pm$ 0.014 & 2.381 $\pm$ 0.009 & $-$0.443 $\pm$ 0.043 & $-$1.160 $\pm$ \nodata & $-$1.340 $\pm$ 0.065 & CIO \\
HR 420         &  7.946 $\pm$ 0.016 & 6.086 $\pm$ 0.009 & ~~~3.409 $\pm$ 0.047 & ~~~2.328 $\pm$ 0.214   & ~~~2.566 $\pm$ 0.045 & P10, 2MASS \\
\enddata
\tablenotetext{a}{Near-infrared references:  2MASS 
  \citep[2MASS All-Sky Point-Source Catalog;][]{skr06},
  CIO \citep[Catalog of Infrared Observations, Ver.\ 
  5.1;][]{gez00}, P10 \citep{pri10}.}
\end{deluxetable*}

\begin{deluxetable*}{lrrlccl} 
\tablenum{8}
\tablecolumns{7}
\tablewidth{0pt}
\tablecaption{Photometry for M giants observed with the IRS}
\label{t.irsphot}
\tablehead{
  \colhead{Target} & \colhead{$B_T$} & \colhead{$V_T$} & 
  \colhead{$J$} & \colhead{$H$} & \colhead{$K$} & \colhead{Near-IR} \\
  \colhead{ } & \colhead{(mag)} & \colhead{(mag)} & 
  \colhead{(mag)} & \colhead{(mag)} & \colhead{(mag)} & 
  \colhead{Reference\tablenotemark{a}}
}
\startdata
HD 13570   &  9.890 $\pm$ 0.024 & 7.992 $\pm$ 0.011 & 4.694 $\pm$ 0.118   & 4.197 $\pm$ 0.076 & 3.775 $\pm$ 0.040 & 2MASS, S04 \\
HD 19554   &  9.693 $\pm$ 0.021 & 7.801 $\pm$ 0.011 & 4.503 $\pm$ 0.107   & 3.603 $\pm$ 0.236 & 3.510 $\pm$ 0.118 & 2MASS, S04 \\
HD 107893  &  9.970 $\pm$ 0.023 & 8.014 $\pm$ 0.011 & 4.686 $\pm$ 0.216   & 3.761 $\pm$ 0.250 & 3.613 $\pm$ 0.010 & 2MASS, S04 \\
\\
HD 17678   & 10.379 $\pm$ 0.032 & 8.388 $\pm$ 0.013 & 4.717 $\pm$ 0.279   & 3.872 $\pm$ 0.262 & 3.652 $\pm$ 0.066 & 2MASS, S04 \\
BD+47 2949 &  9.895 $\pm$ 0.024 & 8.017 $\pm$ 0.012 & 4.810 $\pm$ 0.254   & 3.850 $\pm$ 0.246 & 3.614 $\pm$ 0.286 & 2MASS \\
HD 206503  & 10.345 $\pm$ 0.024 & 8.328 $\pm$ 0.010 & 4.924 $\pm$ 0.037   & 4.033 $\pm$ 0.210 & 3.870 $\pm$ 0.036 & 2MASS \\
\\
HD 122755  & 10.105 $\pm$ 0.027 & 8.132 $\pm$ 0.012 & 4.653 $\pm$ 0.236   & 3.600 $\pm$ 0.216 & 3.385 $\pm$ 0.224 & 2MASS \\
HD 177643  & 10.561 $\pm$ 0.033 & 8.642 $\pm$ 0.013 & 4.972 $\pm$ \nodata & 3.800 $\pm$ 0.198 & 3.820 $\pm$ 0.036 & 2MASS \\
HD 189246  & 10.151 $\pm$ 0.032 & 8.243 $\pm$ 0.014 & 4.623 $\pm$ 0.320   & 3.560 $\pm$ 0.228 & 3.645 $\pm$ 0.322 & 2MASS, S04 \\
\\
HD 26231   & 11.060 $\pm$ 0.047 & 9.093 $\pm$ 0.016 & 4.582 $\pm$ 0.306   & 3.611 $\pm$ 0.254 & 3.451 $\pm$ 0.282 & 2MASS \\
HD 127693  & 11.210 $\pm$ 0.057 & 9.350 $\pm$ 0.019 & 5.171 $\pm$ 0.020   & 4.056 $\pm$ 0.254 & 4.017 $\pm$ 0.036 & 2MASS \\
HD 223306  & 11.076 $\pm$ 0.043 & 9.342 $\pm$ 0.017 & 5.254 $\pm$ 0.021   & 3.980 $\pm$ 0.236 & 3.838 $\pm$ 0.232 & 2MASS \\
\\
HD 17766   & 11.261 $\pm$ 0.055 & 9.242 $\pm$ 0.017 & 5.125 $\pm$ 0.023   & 4.365 $\pm$ 0.076 & 4.107 $\pm$ 0.262 & 2MASS \\
HD 32832   & 10.915 $\pm$ 0.049 & 8.995 $\pm$ 0.017 & 4.884 $\pm$ 0.037   & 4.190 $\pm$ 0.180 & 3.926 $\pm$ 0.258 & 2MASS \\
HD 46396   & 10.442 $\pm$ 0.033 & 8.617 $\pm$ 0.014 & 4.368 $\pm$ 0.254   & 3.472 $\pm$ 0.230 & 3.215 $\pm$ 0.061 & 2MASS, S04 \\
\\
HD 68422   & 11.135 $\pm$ 0.057 & 9.268 $\pm$ 0.020 & 4.421 $\pm$ 0.258   & 3.407 $\pm$ 0.254 & 3.117 $\pm$ 0.268 & 2MASS \\
HD 74584   & 10.380 $\pm$ 0.030 & 8.560 $\pm$ 0.013 & 4.793 $\pm$ 0.254   & 3.670 $\pm$ 0.264 & 3.475 $\pm$ 0.282 & 2MASS \\
HD 76386   & 10.088 $\pm$ 0.028 & 8.186 $\pm$ 0.013 & 4.116 $\pm$ 0.201   & 3.281 $\pm$ 0.238 & 2.912 $\pm$ 0.045 & 2MASS, S04 \\
\\
HD 8680    & 10.944 $\pm$ 0.038 & 9.147 $\pm$ 0.014 & 5.187 $\pm$ 0.037   & 4.201 $\pm$ 0.070 & 4.027 $\pm$ 0.007 & 2MASS, CIO \\
BD+44 2199 & 11.351 $\pm$ 0.058 & 9.818 $\pm$ 0.024 & 4.009 $\pm$ 0.161   & 3.204 $\pm$ 0.194 & 2.759 $\pm$ 0.092 & 2MASS, S04 \\
\enddata
\tablenotetext{a}{Near-infrared references: 2MASS 
  \citep[2MASS All-Sky Point-Source Catalog;][]{skr06},
  CIO \citep[Catalog of Infrared Observations, Ver.\ 
  5.1;][]{gez00}, S04 \citep{smi04}.}
\end{deluxetable*}

To determine the $B-V$ and $V-K$ colors of the larger sample,
we searched for optical and near-infrared photometry 
similarly to the SWS sample.  Optical photometry in the 
Johnson system is less homogeneous and less complete than 
for the SWS sample, so we relied on Tycho magnitudes and the
conversions determined above.  The bright IRS standards are
fainter than the SWS sample, forcing us to rely more on the 
DIRBE photometry \citep{pri10}.  The IRS M giants are even
fainter, and for these we used the calibration of the stellar
DIRBE photometry by \cite{smi04}, which reaches to fainter
magnitudes.  Generally, though, the near-infrared photometry 
is less reliable than for the SWS sample. {\it This lack of 
reliable $JHK$ photometry for reasonably bright sources is 
one of the outstanding shortcomings in existing catalogs!}

\section*{Appendix 2.  On-line spectroscopy}

The spectra presented here are available on-line from the
Infrared Science Archive (IRSA) and VizieR.  They are
organized as simple tables, with columns consisting of
wavelength, flux density, uncertainty in flux density, and
an integer identifying each spectral segment uniquely (1
for SL1, 2 for SL2, 4 for LL1, and 5 for LL2).  IRSA also
provides the data in spectral FITS format, which is identical
to the simple table format, except the rows and columns are
saved as though they were a two-dimensional image.  These
spectra are also available from the first author's website.

\end{document}